\documentclass[aps,prx,onecolumn,showpacs,groupedaddress,superscriptaddress,nofootinbib,longbibliography]{revtex4-2}

\usepackage{mathrsfs}
\usepackage{natbib,hyperref}
\setcitestyle{square,sort&compress,comma,numbers}
\hypersetup{colorlinks=true, citecolor=blue, urlcolor=blue, linkcolor=blue}
\usepackage[english]{babel}
\usepackage[utf8]{inputenc}
\usepackage{graphicx}
\usepackage[caption=false]{subfig}
\usepackage{amsmath}
\usepackage{amsfonts}
\usepackage{amssymb}
\usepackage{easyReview}
\usepackage{xcolor}
\usepackage{siunitx}
\usepackage{color,soul}
\setulcolor{red}\soulregister\cite7
\soulregister\ref7
\soulregister\pageref7
\soulregister\unit7

\usepackage{pdfpages} 
\usepackage{pgffor} 

\makeatletter
\AtBeginDocument{\let\LS@rot\@undefined}
\makeatother

\def\supplementfilename{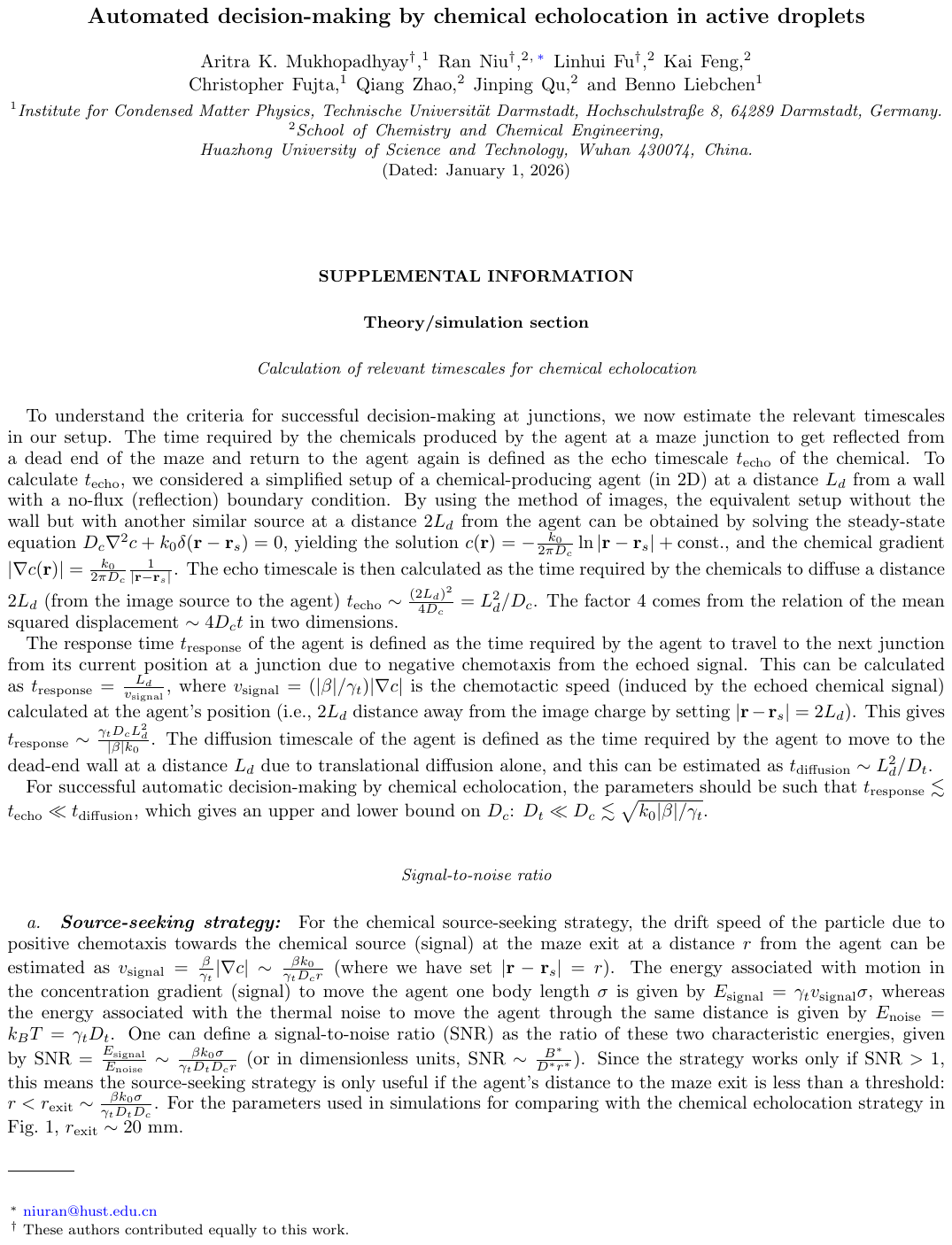}

\pdfximage{\supplementfilename}
\def\numbersupplementpages{\the\pdflastximagepages}

\newif\ifarXiv
\arXivtrue 

\begin{document}
\title{Automated decision-making by chemical echolocation in active droplets}
\author{Aritra K. Mukhopadhyay$^\dagger$}
\affiliation{Institute for Condensed Matter Physics, Technische Universit{\"a}t Darmstadt, Hochschulstraße 8, 64289 Darmstadt, Germany.}
\author{Ran Niu$^\dagger$}
\email{niuran@hust.edu.cn}
\affiliation{School of Chemistry and Chemical Engineering, Huazhong University of Science and Technology, Wuhan 430074, China.}
\author{Linhui Fu$^\dagger$}
\affiliation{School of Chemistry and Chemical Engineering, Huazhong University of Science and Technology, Wuhan 430074, China.}
\author{Kai Feng}
\affiliation{School of Chemistry and Chemical Engineering, Huazhong University of Science and Technology, Wuhan 430074, China.}
\author{Christopher Fujta}
\affiliation{Institute for Condensed Matter Physics, Technische Universit{\"a}t Darmstadt, Hochschulstraße 8, 64289 Darmstadt, Germany.}
\author{Qiang Zhao}
\affiliation{School of Chemistry and Chemical Engineering, Huazhong University of Science and Technology, Wuhan 430074, China.}
\author{Jinping Qu}
\affiliation{School of Chemistry and Chemical Engineering, Huazhong University of Science and Technology, Wuhan 430074, China.}
\author{Benno Liebchen}
\affiliation{Institute for Condensed Matter Physics, Technische Universit{\"a}t Darmstadt, Hochschulstraße 8, 64289 Darmstadt, Germany.}


\date{\today}

\begin{abstract}
Motile microorganisms, like bacteria and algae, unify abilities like self-propulsion, autonomous navigation, and decision-making on the micron scale. While recent breakthroughs have led to the creation of synthetic microswimmers and nanoagents that can also self-propel, they still lack the functionality and sophistication of their biological counterparts. This study pioneers a mechanism enabling synthetic agents to autonomously navigate and make decisions, allowing them to solve mazes and transport cargo through complex environments without requiring external cues or guidance. The mechanism exploits chemo-hydrodynamic signals, produced by agents like active droplets or colloids, to remotely sense and respond to their environment - similar to echolocation. Our research paves the way for endowing autonomous, motile synthetic agents with functionalities that have been so far exclusive to biological organisms.
\end{abstract}


\maketitle
\def\thefootnote{\textdagger}\footnotetext{These authors contributed equally to this work.}\def\thefootnote{\arabic{footnote}}

\section{Introduction}
While nature has already proven the possibility of realizing functional microswimmers, we are not yet able to create them synthetically. For instance, motile microorganisms like bacteria, algae, or slime molds can autonomously move, make navigational decisions, and solve mazes without requiring external cues or guidance \cite{Elgeti_RPP_Physics_Microswimmers_2015, Zhang_PNAS_Collective_Motion_2010, DiLeonardo_PNAS_Bacterial_Ratchet_2010, Rombouts_NC_Multiscale_Dynamic_2023,  Phan_PRX_Bacterial_Route_2020, Dussutour_PNAS_Amoeboid_Organism_2010, Kramar_PNAS_Encoding_Memory_2021, Tweedy_S_Seeing_Corners_2020, Nakagaki_N_Mazesolving_Amoeboid_2000, Raikwar_PRL_Phototactic_DecisionMaking_2025}. Synthetic agents with comparable abilities either depend on large electronic components such as sensors, processors, and actuators \cite{Bandari_NE_Flexible_Microsystem_2020, Mijalkov_PRX_Engineering_Sensorial_2016a} or are not autonomous and depend on external feedback control systems \cite{Smart_S_Magnetically_Programmed_2024, FernandezRodriguez_NC_Feedbackcontrolled_Active_2020, Lavergne_S_Group_Formation_2019, Wang_NC_Spontaneous_Vortex_2023, Fang_AS_DataDriven_Intelligent_2023}. However, recent breakthroughs in physics, chemistry, and nanotechnology have led to a new generation of synthetic agents, called microswimmers, that can self-propel at the micro- and the nanoscale \cite{Paxton_C-EJ_Motility_Catalytic_2005, Bechinger_RMP_Active_Particles_2016, Wang_JACS_Open_Questions_2023}. These agents break the symmetry in their environment, e.g., by catalyzing certain chemical reactions on the part of their surface, which generate flows in the surrounding liquid medium, propelling the agents forward \cite{Golestanian_NJP_Designing_Phoretic_2007, Howse_PRL_SelfMotile_Colloidal_2007}. These agents are now well-established, and their properties have been extensively explored within the field of active matter \cite{Jahanshahi_CP_Realization_Motilitytrap_2020, Kaewsaneha_AAMI_Janus_Colloidal_2013, Oh_NC_Colloidal_Fibers_2019, Zhang_NC_Janus_Particles_2023, DeCorato_PRL_SelfPropulsion_Active_2020, Ziepke_NC_Multiscale_Organization_2022}. For instance, they hold promise for transporting nanocargo and enabling targeted drug delivery to cancer cells \cite{Zhang_SR_Dualresponsive_Biohybrid_2021, Chen_JACS_Deep_Penetration_2021}. However, despite their revolutionary character as engines, synthetic (micro)swimmers do not yet reach the sophistication and functionality of their biological counterparts. In particular, synthetic (micro)swimmers so far lack the ability to autonomously make navigational decisions and to solve mazes without requiring external cues or guidance. If synthetic microswimmers could be endowed with navigation abilities similar to those used by bacteria to locate food, it would open the door towards radically new applications, from environmental remediation \cite{Gao_AN_Environmental_Impact_2014} to the design of functional active materials \cite{Tsang_AIS_Roads_Smart_2020}, meeting the central demands of 21st-century society.

\begin{figure*}
    \begin{center}
        \includegraphics[width=1\textwidth]{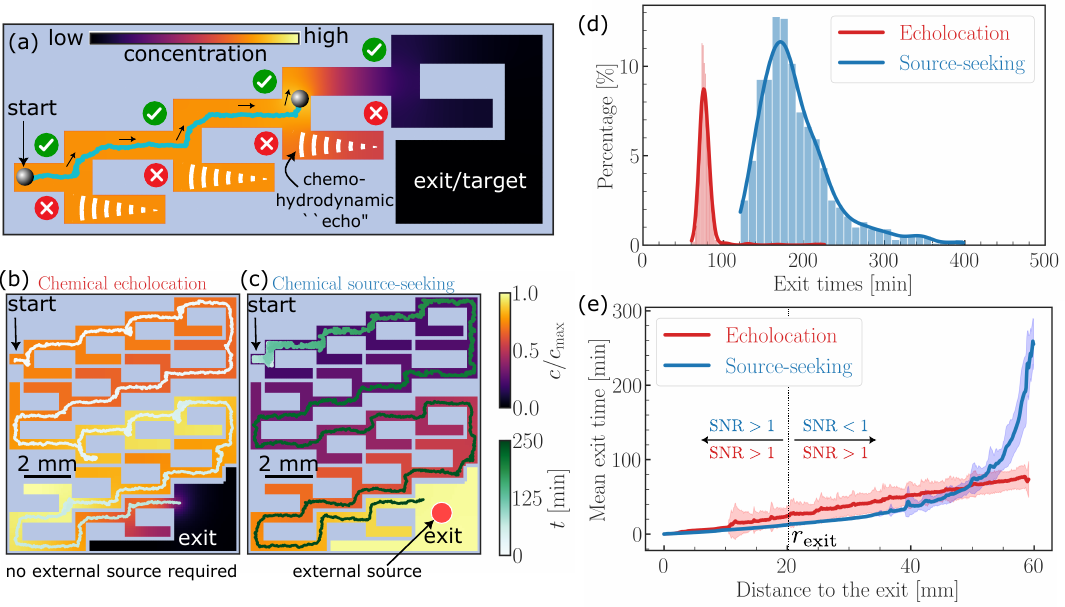}
        \caption{\textbf{Automated decision-making by chemical `echolocation'.} (a) Schematic showing the idea of chemical echolocation. The agent (grey sphere), e.g., a droplet swimmer or an autophoretic colloid, creates a chemical (or chemo-hydrodynamic) signal (background color) that is ‘echoed’ by dead ends in a maze and effectively repels the agent based on physical forces. The agent consistently makes correct navigational decisions (green ticks) toward the target (open end/reservoir) without requiring guidance from a source or external controls. (b) Simulated trajectory (white-green line, color indicates time $t$) showing navigation by the chemical echolocation strategy. (c) Simulated trajectory (white-green line) of a source-seeking agent solving a maze by sensing a chemical gradient created by an external chemical source (red dot) at the exit. Background colors in (b, c) show the normalized chemical concentration $c/c_{\text{max}}$. (d) Histogram of exit times for random initial conditions near `start' in (b) and (c). (e) Average exit time as a function of the distances to the `exit' region in (b) and (c). The signal-to-noise ratio (SNR) is everywhere greater than 1 for the echolocation strategy but smaller than 1 for the source-seeking strategy at large distances from the exit, making it ineffective in large mazes. Parameters: $D^*=10^2$, $B^*=2\times 10^4$, $M^*=0$.}
        \label{fig1}
    \end{center}
\end{figure*}

In the present work, we present a generic mechanism that enables fully synthetic agents to make autonomous navigational decisions and solve complex mazes - a capability previously exclusive to biological systems \cite{Tweedy_S_Seeing_Corners_2020}. We call this mechanism chemical echolocation. It leverages the fact that self-propelled agents, such as autophoretic colloids, droplet swimmers, and ion-exchange-driven modular swimmers, generate chemical concentration gradients and hydrodynamic flow fields in their environment, which they use for propulsion. These fields decay slowly in space and are echoed by (remote) walls, barriers, and other agents. The echo carries information about the agents' surroundings and transmits it back to the agent. Importantly, unlike living organisms, the agent does not require complex information processing to respond to the echo; instead, the echo elicits a direct physical response that steers the agent in a specific direction. For example, when certain droplet swimmers are placed in a complex maze, we ﬁnd that their self-generated chemo-hydrodynamic signals enable them to automatically avoid dead ends and systematically navigate toward the maze's exit, thereby solving the maze autonomously (Fig.~\ref{fig1}a). This work exemplifies how synthetic agents can solve mazes without any external guidance, providing a robust alternative to traditional chemical source-seeking strategies, particularly in large mazes where concentration gradients due to an external source may be weak or absent, and showcasing a fundamentally new approach to micro-scale navigation.

We emphasize that this mechanism fundamentally differs from previous studies involving droplet swimmers and active colloids, which rely on positive chemotaxis (chemoattraction) to solve mazes \cite{Jin_P_Chemotaxis_Autochemotaxis_2017, Gandhi_PRF_Decisionmaking_Tjunction_2020, Lagzi_JACS_Maze_Solving_2010, Pavel_AIS_Autonomous_Chemotactic_2021, Lovass_RA_Maze_Solving_2015, Suzuno_L_Maze_Solving_2014}. In those studies, an external chemical signal is required to guide the agent along a concentration gradient, and without such a gradient, the agent would be unable to locate its target. In contrast, our new strategy \textit{(i) operates effectively even in the complete absence of a chemical source at the maze exit (see Figs.~\ref{fig1}a-c). (ii) Secondly, our approach remains efficient even over very large distances to the exit (see Figs.~\ref{fig1}e)}. Overall, the present work is the first to show that fully synthetic swimmers can exhibit automated decision-making, enabling them to autonomously navigate and solve complex mazes similar to biological agents like Dictyostelium cells \cite{Tweedy_S_Seeing_Corners_2020, Tweedy_FCDB_SelfGenerated_Gradients_2020, Insall_TiCB_Steering_Yourself_2022}, but without requiring their complex information processing machinery. This mechanism is effective at multiple scales, from the cm-scale down to the micro-scale, and can serve as a central ingredient for creating next-generation synthetic agents that can autonomously navigate and deliver drugs and other cargo without requiring any external guidance.

\section{Model} To demonstrate autonomous decision-making by chemical echolocation, we first introduce a minimal model. Consider an agent that produces certain chemicals, such as a chemically active isotropic colloidal particle that catalyzes a certain reaction in its environment \cite{Bechinger_RMP_Active_Particles_2016, Liebchen_ACR_Synthetic_Chemotaxis_2018}, a droplet releasing chemicals from an internal reservoir \cite{Feng_AS_SelfSolidifying_Active_2023}, or a droplet that transforms empty micelles into oil-filled micelles \cite{Jin_P_Chemotaxis_Autochemotaxis_2017, Hokmabad_PRX_Emergence_Bimodal_2021}. These agents are known to exhibit synthetic chemotaxis, responding to gradients in the same chemical fields they generate. This response arises from asymmetric surface flows — diffusioosmotic flows for colloids \cite{Liebchen_ACR_Synthetic_Chemotaxis_2018, Deng_COiC&IS_Active_Colloids_2022, Liebchen_JPCM_Interactions_Active_2021, Zottl_ARCMP_Modeling_Active_2023} and Marangoni flows for droplets \cite{Dwivedi_COiC&IS_Selfpropelled_Swimming_2022, Maass_ARCMP_Swimming_Droplets_2016, Michelin_ARFM_SelfPropulsion_Chemically_2023}. Hence, we model the agent as being able to sense the gradient of a chemical concentration field $c(\textbf{r},t)$ at their position $\textbf{r}_p$ and respond to it with a sensitivity parameter $\beta$. Depending on the sign of $\beta$, the agent moves either toward the gradient (chemoattraction, $\beta>0$) or away from it (chemorepulsion, $\beta<0$). The position $\textbf{r}_p$ of the agent evolves according to the Langevin equation
\begin{equation}
    m \ddot{\textbf{r}}_p (t) + \gamma_t \dot{\textbf{r}}_p (t) = \beta {\nabla} c(\textbf{r}_p,t) + \gamma_t\sqrt{2 D_t}\ \pmb{\eta}(t). \label{part_eqn_dimful}
\end{equation}
Here $m$, $\gamma_t$, and $D_t$ represent the mass, translational dissipation coefficient, and effective diffusion coefficient of the agent, respectively. The term $\beta {\nabla} c(\textbf{r}_p,t)$ models the agent's response to the local chemical gradient. $\pmb{\eta}(t)$ denotes Gaussian white noise of zero mean and unit variance, modeling stochastic fluctuations in the agent's environment. The chemical field $\dot{c} (\textbf{r},t)$ evolves according to the following time-dependent diffusion equation,
\begin{equation}
    \dot{c} (\textbf{r},t) = D_c \nabla^2 c(\textbf{r},t) + k_s(t) \delta \left(\textbf{r}-\textbf{r}_s \right) \label{ch_eqn_dimful}
\end{equation}
where $D_c$ is the effective diffusion coefficient of the relevant chemical species. Besides bare diffusion, this term effectively accounts for the net effect of the advection of the chemicals due to the flow field that is induced by the agent (see Supplemental Material). The second term on the right-hand side of Eq.~\ref{ch_eqn_dimful} accounts for the production of chemical signals at a rate $k_s(t)$ from a point source located at $\textbf{r}_s$. For the conventional source-seeking strategy, a fixed source is placed at the exit of a maze \cite{Jin_P_Chemotaxis_Autochemotaxis_2017, Gandhi_PRF_Decisionmaking_Tjunction_2020, Lagzi_JACS_Maze_Solving_2010, Pavel_AIS_Autonomous_Chemotactic_2021, Lovass_RA_Maze_Solving_2015, Suzuno_L_Maze_Solving_2014, Kim_B_Investigations_Design_2008, Reynolds_PRE_Mazesolving_Chemotaxis_2010, Salek_NC_Bacterial_Chemotaxis_2019}. In contrast, here, we treat the agent itself as a source, such that the source location coincides with the position of the moving agent, i.e., $\textbf{r}_s = \textbf{r}_p(t)$. To solve these equations, we have developed a numerical hybrid particle-continuum solver taking into account complex boundary conditions due to the maze and carefully accounting for the conservation of the total chemical concentration for $k_s=0$ (see Supplemental Material).

\section{Automated decision-making in mazes} 
Unlike the chemical source-seeking strategy ($\beta>0$), where the agent continuously moves towards an imposed source \cite{Jin_P_Chemotaxis_Autochemotaxis_2017, Gandhi_PRF_Decisionmaking_Tjunction_2020, Lagzi_JACS_Maze_Solving_2010, Pavel_AIS_Autonomous_Chemotactic_2021, Lovass_RA_Maze_Solving_2015, Suzuno_L_Maze_Solving_2014}, we now consider the case where $\beta<0$ and $\textbf{r}_s = \textbf{r}_p(t)$, such that the agent moves away from its self-produced chemicals and is not guided by any external sources. To explore how the agent navigates through a maze, let us choose a constant production rate $k_s(t)=k_0$ and expose it to a maze made of consecutive Y-junctions (Fig.~\ref{fig1}b). At each junction, the agent has to make a `decision' between a dead end and the correct path leading to the maze exit (an outlet into an open region or reservoir). We initialize the agent at the `start' position (Fig.~\ref{fig1}b). Fig.~\ref{fig1}b and Supplemental Movie M1 show the dynamics of a representative agent trajectory. Initially, the agent moves toward the first junction and quickly reaches it. At the first junction, it briefly turns towards the wrong path (a dead end) but quickly corrects its direction to take the correct path towards the exit (see also Supplemental Movie M1). Similarly, at subsequent junctions, the agent either directly chooses the correct path or briefly enters the wrong path before turning around. That is, remarkably, at each Y-junction, the agent systematically makes the correct navigational decision to progress toward the target. Since this behavior arises solely from the agent’s response to physical forces acting on its surface \cite{Liebchen_ACR_Synthetic_Chemotaxis_2018}, without any external control, we refer to this capability as ‘automated decision-making.’ This mechanism demonstrates how a purely physical mechanism can enable synthetic microswimmers to exhibit sophisticated, autonomous behaviors akin to those seen in biological systems \cite{Tweedy_S_Seeing_Corners_2020, Tweedy_FCDB_SelfGenerated_Gradients_2020, Insall_TiCB_Steering_Yourself_2022}.

\begin{figure*}[t!]
    \begin{center}
        \includegraphics[width=1\textwidth]{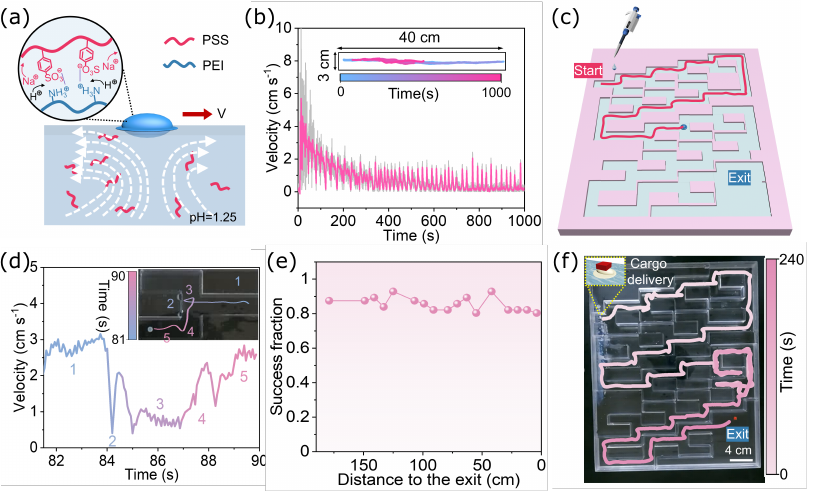}
        \caption{\textbf{Experimental realization of automated decision-making in droplet swimmers.} (a) Schematic of the droplet showing the asymmetric distribution of the released PSS, leading to spontaneous symmetry breaking and self-propulsion. (b) Speed of the droplet in a rectangular channel. The inset shows the corresponding trajectory in this confined channel. (c) Schematic of the droplet motion in a large complex maze. (d) Variation of the droplet velocity at a junction. The inset shows the trajectory of the droplet. (e) Success fraction (SF) of the droplet at each junction on the first attempt versus distance from the exit. (f) Autonomous large cargo delivery without requiring external control.}
        \label{fig2}
    \end{center}
\end{figure*} 

\section{Chemical echolocation} Let us now analyze the mechanism that makes automated decision-making possible for the agent in detail. The chemical produced by the agent diffuses through the maze and is reflected by the walls or maze boundaries. At the junctions, where the path bifurcates into a dead end and an open path, chemicals accumulate in the dead arms, creating a higher concentration in the dead arm compared to the open arm,  which essentially represents a `chemical echo' (Fig.~\ref{fig1}a). For very low chemotactic sensitivity (i.e., low value of $|\beta|$), the agent's motion is diffusion-dominated, and it enters either arm with near-equal likelihood. However, if the sensitivity parameter is sufficiently large, the agent can detect this echoed signal at the junction, effectively identifying the dead end from a distance, akin to echolocation used by bats for navigation or ships for measuring water depths \cite{Moss_FBN_Probing_Natural_2010}, without physically entering the dead arm. The agent then moves away from the high chemical concentration in the dead end, turning towards the open path (Fig.\ref{fig1}a), thereby making the correct navigational decision to progress toward the exit. This shows how the interplay between chemical production, diffusion, and reflection creates a chemical echo from which the agent automatically steers, using synthetic chemotaxis, to systematically make the correct navigational decisions to solve mazes.

What are the criteria for such automated decision-making? There are three characteristic timescales for the agent. 1) The echo timescale, i.e., the time for the chemical signal to travel from the agent's position, when it is at the junction, to the dead end and back. This can be estimated as $t_{\text{echo}}\sim L_d^2/D_c$ (see Supplemental Material), where $L_d$ is the length of the dead arm. 2) The response time of the agent, defined as the time it takes to travel to the next junction due to negative chemotaxis, is estimated as $t_{\text{response}} \sim \frac{\gamma_t D_c L_d^2}{|\beta| k_0}$ (see Supplemental Material). This timescale represents the window in which the agent must react to the echoed signal before overshooting into a dead end. 3) The diffusion timescale, which is the time required for the agent to traverse the dead-arm length due to translational diffusion alone, can be estimated as $t_{\text{diffusion}} \sim L_d^2/D_t$. For chemical echolocation to work, the agent's response time should be less than the echo timescale of the chemical signal, and both these timescales should be significantly lower than the agent's diffusion timescale, i.e., $t_{\text{response}} \lesssim t_{\text{echo}} \ll t_{\text{diffusion}}$. This imposes an upper and lower bound for $D_c$: $D_t \ll D_c \lesssim \sqrt{k_0|\beta|/\gamma_t}$. This criterion is met in our simulations underlying Fig.~\ref{fig1} and in typical droplet swimmers, for which we present experiments further below.


\section{Advantages of chemical echolocation over source-seeking strategy} How does this new mechanism for automated decision-making compare to the well-established chemical source-seeking strategy, in which agents follow the gradient due to an external source (Fig.~\ref{fig1}c) \cite{Jin_P_Chemotaxis_Autochemotaxis_2017, Gandhi_PRF_Decisionmaking_Tjunction_2020, Lagzi_JACS_Maze_Solving_2010, Pavel_AIS_Autonomous_Chemotactic_2021, Lovass_RA_Maze_Solving_2015, Suzuno_L_Maze_Solving_2014, Kim_B_Investigations_Design_2008, Reynolds_PRE_Mazesolving_Chemotaxis_2010, Salek_NC_Bacterial_Chemotaxis_2019, Adamatzky_SM_Liquid_Metal_2020}? To conduct this comparison, we simulated chemoattractive agents (i.e., $\beta>0$) in the same maze, but with a chemical source placed at a position $\textbf{r}_s$ near the exit. This source produces chemicals at a rate $k_0$ (Fig.~\ref{fig1}c). While this mechanism is effective for small and moderately sized mazes, for mazes beyond a certain size (Fig.~\ref{fig1}c,e), it becomes ineffective. As shown in Fig.~\ref{fig1}c, a typical agent in this scenario spends a very long time near its starting position, showing diffusive motion before reaching the first junction (see Supplemental Movie M2). It continues to move diffusively through the initial junctions, only speeding up as it approaches the source at the exit. This behavior is in stark contrast to chemical echolocation, where the average speed of the agent is relatively constant regardless of its distance from the exit. As a consequence, for sufficiently large mazes, an echolocating agent reaches the exit much faster than a source-seeking agent, as evidenced by Figs.~\ref{fig1}e and the time colorbars in Figs.~\ref{fig1}b,c. 

Why does the source-seeking agent take much longer to reach the exit of a large maze? In two dimensions, the gradient $|\nabla c|$ decays as $\sim 1/r$ with increasing distance $r$ from the source. This decay results in a diminished signal at larger distances from the exit, causing the agent to exhibit prolonged diffusive dynamics dominated by translational noise during the initial stages of its journey (Figs.~\ref{fig1}c). Critically, the source-seeking strategy becomes ineffective when the SNR, defined as the ratio between the energy cost of the signal and thermal noise, respectively, to move the agent one body length, drops below 1. This occurs when the agent is beyond a threshold distance $r_\text{exit}\sim\frac{\beta k_0 \sigma}{\gamma_t D_t D_c}\approx 20 \unit{mm}$ from the maze exit for typical parameters for mm-sized droplet swimmers, as used in our simulations (see Figs.~\ref{fig1}e). In stark contrast, the chemical echolocation strategy maintains a constant SNR $\approx 10$ independent of distance, enabling more reliable navigation through complex mazes (see Supplemental Material for detailed derivation).

To statistically compare the performance of the two strategies, we analyzed $10^3$ agent trajectories for each strategy. For each trajectory, we simulated the dynamics of the agent and the corresponding chemical concentration field. In each simulation, the agent was initialized at a random position near the maze entrance, and its exit time was recorded. Our analysis reveals that for the maze considered in Fig.~\ref{fig1}b,c the average exit time of the chemical echolocating agents ($\approx 78.2 \pm 0.4$ \unit{\min}) was substantially shorter than that of chemical source-seeking agents ($\approx 191 \pm 1$ \unit{\min}) (Fig.\ref{fig1}d). Crucially, in general, the average exit time increases almost exponentially with the distance from the exit for the source-seeking agent, whereas for echolocating agents, the increase is approximately linear  (Fig.\ref{fig1}e). This disparity highlights a critical limitation of the source-seeking strategy: while it performs well in small to medium-sized mazes, for instance, in \cite{Jin_P_Chemotaxis_Autochemotaxis_2017}, where the shortest path was $\sim 6$ \unit{\mm} ($\approx 60$ times the agent diameter), agents exited in $\sim 5$ min. It becomes increasingly inefficient in larger mazes. In our simulations (using the same agent parameters as \cite{Jin_P_Chemotaxis_Autochemotaxis_2017}; see Supplemental Material), where the shortest path is $\sim 60$ \unit{\mm} (10 times larger), the average exit time for source-seeking agents surged to $191$ min (40 times longer). In contrast, chemically echolocating agents navigated the same large maze significantly faster, exiting in $\approx 78$ min, by making accurate decisions at each junction without relying on a chemical source at the exit. It is important to note that while echolocation enables effective maze traversal and obstacle avoidance, traditional chemotaxis is particularly effective for final target acquisition once the agent is sufficiently close to the target (such that the SNR is well above 1.)

\section{Realization of automated decision-making in droplet swimmers} We now present an experimental realization of the chemical echolocation strategy using aqueous droplets composed of a mixture of two polymers, poly(ethyleneimine) (PEI) and poly(sodium 4-styrenesulfonate) (PSS). When placed in acidic water, these droplets solidify from their edges, forming a porous outer shell that gradually releases PSS (see Supplemental Fig.~S2) into the surrounding aqueous solution (Fig.~\ref{fig2}a) \cite{Feng_AS_SelfSolidifying_Active_2023}. The released PSS lowers the local surface tension of water and generates Marangoni flows at the droplet interface. These flows propel the droplet from regions of higher to lower polymer concentrations. As a result, the droplets effectively exhibit negative chemotaxis, avoiding areas of high PSS concentration. This behavior fulfills the key requirements for chemical echolocation, as the droplets act as autonomous agents that both produce and respond to chemical gradients, analogously as in our model and simulations. In confined channels of width $\approx$ 8 droplet diameters, these droplets perform directed motion until they reach the boundary and are bounced back (Fig.~\ref{fig2}b). This behavior demonstrates the droplet’s ability to navigate confined environments, a critical feature for implementing chemical echolocation in complex geometries.

\begin{figure*}[t!]
    \begin{center}
        \includegraphics[width=1\textwidth]{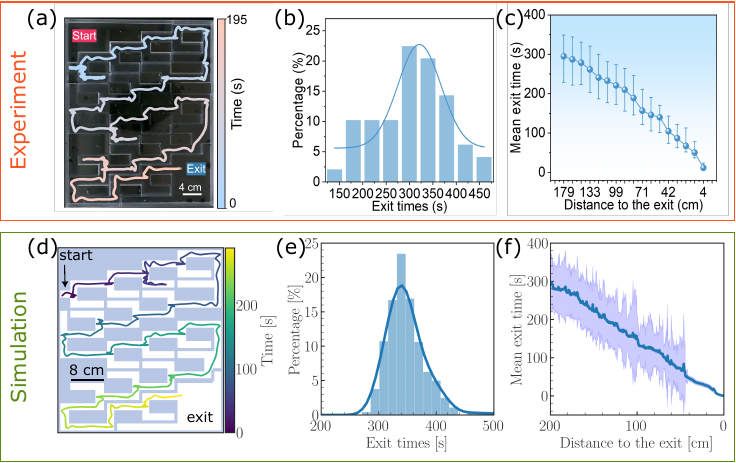}
        \caption{\textbf{Comparison of experiments and simulations.} Typical trajectory of a droplet in the maze from (a) experiments and (d) simulations, with time shown in colorbars. Histogram of exit times of droplet trajectories in (b) experiments and (e) simulations. The solid lines are Gaussian fits to the data. Mean exit time of the droplet from different distances to the exit point in (c) experiments and (f) simulations. Error bars in c are first-to-third quartile of the distributions. Parameters: $D^*=2\times 10^2$, $B^*=-8\times 10^4$, and $M^*=10^{-1}$ (for panels d,e,f).}
        \label{fig3}
    \end{center}
\end{figure*} 

We designed a large maze with the same geometry as in Fig.~\ref{fig1}b, featuring 17 junctions, and initially filled it with an acidic water solution of pH 1.25 (Fig.~\ref{fig2}c). A droplet ($\sim 4$ mm diameter) of the PEI/PSS solution was placed at the maze entrance from a height of $\sim$ 1 cm above the air-water interface. Upon contact, the droplet began to self-propel while gradually solidifying from the outer edge inward \cite{Feng_AS_SelfSolidifying_Active_2023}. Inside the maze, it moves at an average speed of $\approx 1$ \unit{cm.s^{-1}} ($\approx 2.5$ body lengths per second) (Fig.~\ref{fig2}d), slowing down as it approaches a junction and moving faster along the open channels. A representative trajectory, recorded with a high-speed camera (see Fig.~\ref{fig3}a and Supplemental Movie M3), shows the droplet's ability to efficiently navigate the maze: it made correct decisions at most junctions, avoided dead ends, and reached the exit in $\approx 195$ s. We repeated the experiment with 60 droplets and analyzed the distribution of exit times (Fig.~\ref{fig3}b), estimating a mean exit time of $\approx 321 \pm 46$ \unit{\second}. Importantly, the mean exit time scales linearly with the distance to the exit (Fig.~\ref{fig3}c), confirming the efficiency of the droplet’s echolocation strategy, as predicted in the previous section. We also analyzed the success fraction at each junction, which is defined as the fraction of trajectories that choose the correct path toward the exit. The average success fraction was $\approx 0.8$ (see Fig.~\ref{fig2}e), indicating that the droplets made correct decisions at most junctions. Importantly, using particle image velocimetry (PIV), we see the hydrodynamic backflow generated from the dead end at the junction, which pushes the droplet to the open end (see Supplemental Material, Fig.~S7), confirming the chemo-hydrodynamic echo mechanism. This autonomous decision-making capability highlights the potential for intelligent cargo delivery in complex pathways, such as blood vessels. To demonstrate this, we loaded a piece of cardboard and a foam sphere of diameter 3 mm as a representative cargo onto the droplet. Leveraging the droplet’s chemical echolocation ability, the cargo was successfully delivered to the target via the shortest path in 240 s and 206 s, respectively (Fig.~\ref{fig2}f and Fig.~S8). To further validate our findings, we performed numerical simulations using the parameters derived from our droplet experiments (see Supplemental Material). The results show strong quantitative and qualitative agreement with the experimental data, underscoring the robustness and reliability of the proposed mechanism (cf. Fig.~\ref{fig3}d-f).

\begin{figure*}[t!]
    \begin{center}
        \includegraphics[width=1\textwidth]{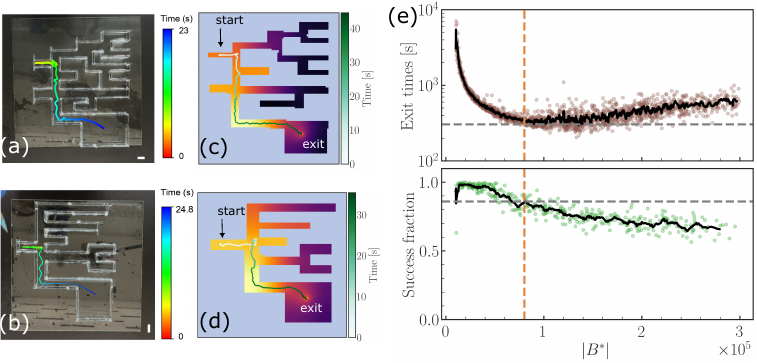}
        \caption{\textbf{Robustness of chemical echolocation for varying maze geometry and parameters.} Representative agent trajectories from (a,b) experiments and (c,d) simulations demonstrate successful navigation through complex mazes with extended dead-end lengths and multiple branches, highlighting the effectiveness of the chemical-echolocation-based decision-making strategy. Parameters for panels (c,d) are the same as in Figs.~\ref{fig3} (d-f). Background colors in (c, d) show the normalized chemical concentration $c/c_{\text{max}}$ when the droplet reaches the exit. (e) Exit times and success fraction of the droplet for the maze shown in Fig.~\ref{fig3}d for different values of the reduced chemotactic sensitivity $B^*$. The black solid lines denote moving averages. The black dashed lines indicate the average values of the exit times (Fig.~\ref{fig3}b) and success fractions (Fig.~\ref{fig2}e) of the droplet from the experiments. The orange dashed line indicates the value of  $B^*=-8\times 10^4$ used in the Figs.~\ref{fig3} (d-f). Parameters: $D^*=2\times 10^2$ and $M^*=10^{-1}$.}
        \label{fig4}
    \end{center}
\end{figure*} 

\section{Robustness against variations of parameters and maze geometry} The chemical echolocation mechanism proves effective even in highly complex mazes featuring significantly longer and more dead-end branches. We designed two intricate maze configurations featuring extended dead-end paths and multiple branches. Our observations demonstrate that the droplet agent can effectively utilize chemical echolocation to avoid dead ends and successfully navigate toward the exit, as illustrated in Fig.~\ref{fig4}a,b (experiments) and Fig.~\ref{fig4}c,d (simulations). Note that the timescales in the simulations are somewhat longer than in experiments. This is largely because the model effectively represents the key effects of chemical echolocation, but not all details, which can matter particularly for maze geometries where chemohydrodynamic signals from different dead ends significantly overlap. The average success fraction is 75-80\% when dead-ends and open ends have comparable lengths, and even if the dead-end is twice as long as the desired path, the success fraction is $\gtrsim  60\% $(see Supplementary Fig.~S9(a)). Similarly, chemical echolocation is also still efficient when the available space within the dead-end arms significantly increases compared to the droplet size (see Supplementary Figs.~S9(b) and S10). Furthermore, experiments varying the droplet size reveal that navigation remains robust across a wide range of channel confinements (channel width to droplet diameter ratio in the range of $\approx$ 5-40), only degrading when the channel becomes too wide ($\gtrsim 50$ times the droplet diameter) to support a directed echo towards the open pathways (see Supplementary Fig.~S11). Further experiments with droplets starting from different positions in the maze confirmed successful navigation to the exit, though performance was more reliable when the starting point was closer to the maze entrance (see Supplementary Fig.~S6(a-e)). In experiments with multiple droplets released simultaneously from different locations, the droplets successfully reached the exit despite occasionally taking incorrect turns at junctions before correcting their path (see Supplementary Fig.~S6f). In such multi-agent scenarios, the overlapping chemical gradients and hydrodynamic wakes can interfere with the echolocation signal, leading to complex collective behaviors that differ from single-agent navigation (see Supplemental Movie M4). This will be explored further in future work. We further investigated droplet navigation efficiency as a function of chemotactic sensitivity (i.e., $\beta$). An optimal navigation mechanism should simultaneously minimize exit time while maximizing success probability at each decision junction. Through systematic simulations with varying dimensionless chemotactic sensitivity ($B^*=\beta k_0/\gamma_t D_t^2$) while holding other parameters constant, we quantified exit times and success fractions using a Uniform Cost Search (UCS) algorithm (see Supplemental Material and Fig.~S3 for details). For small $|B^*|$, the chemotactic response is weak compared to translational diffusion ($t_{\mathrm{response}} \gg t_{\mathrm{echo}}$), so navigational decisions at junctions are dominated by random motion, leading to long exit times (Fig.~\ref{fig4}e). As $|B^*|$ approaches $\sim 4\times 10^4$, the response time becomes shorter than the echo timescale, enabling reliable detection of chemical echoes and optimal decision-making at junctions. This decreases the exit times sharply and increases the success fraction to approximately unity. However, beyond this point, the success fraction begins to decline, with only a marginal further reduction in exit times. Notably, within the range $8\times 10^4<|B^*|<1.4\times 10^5$, exit times remain nearly constant, indicating an optimal range for efficient maze-solving via chemical echolocation. For very large values of $|B^*|$, the droplet's response becomes excessively strong, leading to irregular motion as it reacts to minor chemical gradients from multiple directions, and not solely to those originating from dead ends. Consequently, this reduces the success fraction and, in cases of extremely large $|B^*|$, increases the exit time. To match the average exit time and success fraction observed in our experiments, we chose $B^*=-8\times 10^4$ in our simulations (Fig.~\ref{fig3}d-f). This shows that automated decision-making by chemical echolocation is highly robust, both concerning parameter variations and changes in the maze geometry.

\section{Key requirements of chemical echolocation and downscaling potential} We identify two key conditions required for effective chemical echolocation based navigation: (i) the chemical diffusivity should be neither extremely fast nor extremely slow, i.e.,  $D_t \ll D_c \lesssim \sqrt{k_0|\beta|/\gamma_t}$ as specified above, and (ii) the target must be situated either at an open end of the maze or within a larger reservoir where chemicals can accumulate without generating strong echoes. Following (i), the mechanism can be remarkably downscaled to microdroplets of size down to $\sim 100$ \unit{\um} diameter and other chemo-hydrodynamically active agents such as Janus colloids with sizes down to $\sim 1$ \unit{\um} (see Supplemental Material). Experimental validation using 1 mm and 500 \unit{\micro m} droplets showed high decision accuracy at maze junctions (Fig.~S4), demonstrating successful chemical echolocation consistent with our predictions. Our analysis shows that other systems, such as bubble-propelled microengines \cite{Solovev_AFM_Magnetic_Control_2010, Sanchez_JACS_Microbots_Swimming_2011, Gao_JACS_Highly_Efficient_2011, Lin_ACIE_BubblePropelled_Janus_2021} or other oil droplets \cite{Dwivedi_JCP_Chemical_Interactions_2025}, can also effectively utilize this navigation strategy (see Supplemental Material Table S1 for a comparison of different setups, regarding their potential to implement chemical echolocation). Further downscaling is limited primarily by the current setup. But when using other setups to realize much smaller droplets, further downscaling should work to some point, beyond which the chemo-hydrodynamic signals are dominated by noise (e.g., at the lower micro- and the nanoscale). To implement the mechanism at the nanoscale, one would need alternate setups which either reduce the agent’s drag coefficient (e.g., by changing the medium \cite{Nosenko_PRR_Active_Janus_2020}) or employ agents that produce chemicals at much higher rates or exhibit enhanced chemical sensitivity to fulfill the criteria (i) above. Looking forward, these properties can be controlled to some extent via surface functionalization in active colloids. Synthetic biology approaches may also enable enhancement of the chemical emission rates and sensing capabilities over a much wider range in artificial cells and other life-like system components in the future. These developments could extend the principle to sub-micron regimes, with potential applications in targeted drug delivery and micro-robotic systems.

\section{Conclusions} Our work introduces a generic mechanism to endow synthetic agents with the ability to autonomously make navigational decisions without requiring any external guidance. This capability has so far been reserved for biological agents and large sensor-processor-actuator systems like electronic robots. By leveraging self-generated chemo-hydrodynamic gradients, we demonstrate that synthetic agents, such as droplet swimmers, can autonomously solve complex environments by using their self-produced concentration fields to remotely sense obstacles (dead ends in a maze) without relying on any external stimuli. The mechanism facilitates fast and accurate decision-making in intricate maze environments and scales favorably with maze size, providing a robust alternative to traditional chemical source-seeking strategies, particularly in large mazes where concentration gradients due to an external source may be weak or absent. Crucially, since it does not require electronic sensor-processor-actuator components, this approach is ideally suited for miniaturization of the agent. In all, this work provides a key ingredient for the realization of future autonomous robotic systems capable of independent navigation and designed to meet the central healthcare and environmental remediation demands of 21st-century society. Future work could extend beyond the effective model used here to resolve the full chemo-hydrodynamics of the droplet within the maze, providing deeper insight into the precise dynamics at maze junctions.


\begin{acknowledgments}
A.K.M. and B.L. acknowledge financial support from the Deutsche Forschungsgemeinschaft (DFG, German Research Foundation) in the framework of the collaborative research center Multiscale Simulation Methods for Soft-Matter Systems (TRR 146) under Project No. 233630050. R.N. thanks the financial support from the National Natural Science Foundation of China (No. 22102059).
\end{acknowledgments}

\section*{Author contributions}
A.K.M., R.N., and L.F. contributed equally to this work. B.L. initiated the project. R.N. and B.L. designed the experiments and simulations, respectively. R.N., L.F., and K.F. performed the experiments and analyzed the data; A.K.M. and C.F. conducted the simulations and analyzed the data. A.K.M. and R.N. created the visualizations and data presentations. A.K.M., B.L., and R.N. participated in writing the manuscript. All authors have discussed the results. All authors approved the final version of the manuscript.

\appendix
\section{Model and simulation method} 
First, to reduce the parameter space, we choose the units of length, time, and concentration as $r_u=\sigma$ (agent diameter), $t_u=\sigma^2/D_t$ (characteristic translational diffusion time of agent), and $c_u=k_0/D_t$ (amount of chemical produced by the agent in time $t_u$), respectively, which gives us the dimensionless equations of motion
\begin{eqnarray}
    \dot{c}^*(\textbf{r}^*,t^*) &=& D^* \nabla^2 c^*(\textbf{r}^*,t^*) + \delta (\textbf{r}^*-\textbf{r}^*_s) \label{Methods_ch_eqn_dimless}\\
    M^*\ddot{\textbf{r}^*_p} + \dot{\textbf{r}^*_p} &=& B^* {\nabla} c^*(\textbf{r}^*_p,t^*) + \sqrt{2}\ \pmb{\eta}(t^*) \label{Methods_part_eqn_dimless}
\end{eqnarray}
Our system is thus characterized by the three dimensionless parameters: the reduced chemical diffusion coefficient $D^*=D_c/D_t$, the reduced particle mass $M^*=mD_t/\gamma_t\sigma^2$, and the reduced chemotactic sensitivity of the particles $B^* =\beta k_0/\gamma_t D_t^2$. For our simulations, we numerically integrate the coupled equations Eq.~\ref{Methods_ch_eqn_dimless} and Eq.~\ref{Methods_part_eqn_dimless} iteratively at the same time:  using a finite-difference scheme on a discretized spatial grid of size $\sigma$ for Eq.~\ref{Methods_ch_eqn_dimless} and using a forward Euler-Maruyama scheme with timestep $dt=10^{-3}$ for Eq.~\ref{Methods_part_eqn_dimless} with no-flux boundary conditions $(\nabla c^*=0)$ for the chemicals at all the maze walls. The two-dimensional $\delta$ function in Eq.~\ref{Methods_ch_eqn_dimless} is approximated by a square grid-point source of strength $1/\sigma^2$. The gradient in Eq.~\ref{Methods_part_eqn_dimless} is calculated at the grid point nearest to the particle coordinate $\textbf{r}^*_p$ using a finite-difference scheme. To prevent the collision of agents with maze walls, we imposed a very short-ranged (cut-off distance of $\sigma$) Weeks-Chandler-Anderson potential between the agent and the maze walls.

\section{Preparation of the PEI/PSS solution} We obtain Polyethylenimine (PEI, branched, linear formula:(CH2CH2NH)n, average molecular weight $\approx$ 70000, 50 wt$\%$ in H2O), HCl from Sinopharm Chemical Reagent Co., Ltd (China). Poly(sodium 4-styrenesulfonate) (PSS, average molecular weight $\approx$ 1000000) was purchased from Sigma-Aldrich (USA). Deionized water was purified by a ULUP water purification system with a minimum resistivity of 18.25 $\unit{m \Omega cm^{-1}}$ and used in all the processes. Typically, 2.06 g PSS powder and 0.43 g PEI solution (50 wt$\%$) were added into 8.88 mL of deionized water and mixed thoroughly on a magnetic stirrer for 6 h to obtain a maize-yellow solution for further use.

\section{Designing the maze} We fabricated the maze channels with different widths made from polymethyl methacrylate (PMMA) that was designed by the groove milling method. The channel was 0.2 cm thick and 2.0 cm deep. All the channels have the same width. The maze includes 17 junctions and a buffer region while keeping the same width of all channels. Mazes of other geometries were fabricated by gluing glass slides of desired length onto a big glass bottom using two-component epoxy glue, and waited for four days before use.

\section{Data, Materials, and Software Availability} The codes for numerical simulations can be found here: https://tudatalib.ulb.tu-darmstadt.de/handle/tudatalib/4934. The supplemental movies can be found here: https://tudatalib.ulb.tu-darmstadt.de/handle/tudatalib/4972.

\bibliographystyle{apsrev_mod}
\bibliography{mybib}

\ifarXiv


\section*{Supplementary material}

# Automated decision-making by chemical echolocation in active droplets

Aritra K. Mukhopadhyay[†],[1] Ran Niu[†],[2, *] Linhui Fu[†],[2] Kai Feng,[2]
Christopher Fujta,[1] Qiang Zhao,[2] Jinping Qu,[2] and Benno Liebchen[1]

[1]*Institute for Condensed Matter Physics, Technische Universität Darmstadt, Hochschulstraße 8, 64289 Darmstadt, Germany.*
[2]*School of Chemistry and Chemical Engineering,
Huazhong University of Science and Technology, Wuhan 430074, China.*
(Dated: January 1, 2026)

## SUPPLEMENTAL INFORMATION

### Theory/simulation section

*Calculation of relevant timescales for chemical echolocation*

To understand the criteria for successful decision-making at junctions, we now estimate the relevant timescales in our setup. The time required by the chemicals produced by the agent at a maze junction to get reflected from a dead end of the maze and return to the agent again is defined as the echo timescale $t_{\text{echo}}$ of the chemical. To calculate $t_{\text{echo}}$, we considered a simplified setup of a chemical-producing agent (in 2D) at a distance $L_d$ from a wall with a no-flux (reflection) boundary condition. By using the method of images, the equivalent setup without the wall but with another similar source at a distance $2L_d$ from the agent can be obtained by solving the steady-state equation $D_c \nabla^2 c + k_0 \delta(\mathbf{r} - \mathbf{r}_s) = 0$, yielding the solution $c(\mathbf{r}) = -\frac{k_0}{2\pi D_c} \ln |\mathbf{r} - \mathbf{r}_s| + \text{const.}$, and the chemical gradient $|\nabla c(\mathbf{r})| = \frac{k_0}{2\pi D_c} \frac{1}{|\mathbf{r} - \mathbf{r}_s|}$. The echo timescale is then calculated as the time required by the chemicals to diffuse a distance $2L_d$ (from the image source to the agent) $t_{\text{echo}} \sim \frac{(2L_d)^2}{4D_c} = L_d^2/D_c$. The factor 4 comes from the relation of the mean squared displacement $\sim 4 D_c t$ in two dimensions.

The response time $t_{\text{response}}$ of the agent is defined as the time required by the agent to travel to the next junction from its current position at a junction due to negative chemotaxis from the echoed signal. This can be calculated as $t_{\text{response}} = \frac{L_d}{v_{\text{signal}}}$, where $v_{\text{signal}} = (|\beta|/\gamma_t)|\nabla c|$ is the chemotactic speed (induced by the echoed chemical signal) calculated at the agent's position (i.e., $2L_d$ distance away from the image charge by setting $|\mathbf{r} - \mathbf{r}_s| = 2L_d$). This gives $t_{\text{response}} \sim \frac{\gamma_t D_c L_d^2}{|\beta| k_0}$. The diffusion timescale of the agent is defined as the time required by the agent to move to the dead-end wall at a distance $L_d$ due to translational diffusion alone, and this can be estimated as $t_{\text{diffusion}} \sim L_d^2/D_t$.

For successful automatic decision-making by chemical echolocation, the parameters should be such that $t_{\text{response}} \lesssim t_{\text{echo}} \ll t_{\text{diffusion}}$, which gives an upper and lower bound on $D_c$: $D_t \ll D_c \lesssim \sqrt{k_0 |\beta|/\gamma_t}$.

*Signal-to-noise ratio*

*a.* **Source-seeking strategy:** For the chemical source-seeking strategy, the drift speed of the particle due to positive chemotaxis towards the chemical source (signal) at the maze exit at a distance $r$ from the agent can be estimated as $v_{\text{signal}} = \frac{\beta}{\gamma_t} |\nabla c| \sim \frac{\beta k_0}{\gamma_t D_c r}$ (where we have set $|\mathbf{r} - \mathbf{r}_s| = r$). The energy associated with motion in the concentration gradient (signal) to move the agent one body length $\sigma$ is given by $E_{\text{signal}} = \gamma_t v_{\text{signal}} \sigma$, whereas the energy associated with the thermal noise to move the agent through the same distance is given by $E_{\text{noise}} = k_B T = \gamma_t D_t$. One can define a signal-to-noise ratio (SNR) as the ratio of these two characteristic energies, given by $\text{SNR} = \frac{E_{\text{signal}}}{E_{\text{noise}}} \sim \frac{\beta k_0 \sigma}{\gamma_t D_t D_c r}$ (or in dimensionless units, $\text{SNR} \sim \frac{B^*}{D^* r^*}$). Since the strategy works only if $\text{SNR} > 1$, this means the source-seeking strategy is only useful if the agent's distance to the maze exit is less than a threshold: $r < r_{\text{exit}} \sim \frac{\beta k_0 \sigma}{\gamma_t D_t D_c}$. For the parameters used in simulations for comparing with the chemical echolocation strategy in Fig. 1, $r_{\text{exit}} \sim 20$ mm.

---

* niuran@hust.edu.cn
† These authors contributed equally to this work.

b. **Echolocation strategy:** In contrast, for the chemical echolocation strategy, the drift speed of the agent induced by the echoed chemical signals from the dead-end is given by $v_{\text{signal}} \sim \frac{\beta k_0}{\gamma_t D_c L_d}$ (i.e., $2L_d$ distance away from the image charge by setting $|\mathbf{r} - \mathbf{r}_s| = 2L_d$, see the previous section for details). Unlike the source-seeking strategy, note that the drift speed is independent of the distance to the maze exit, but depends on the distance to the nearest wall $L_d$, which is $\sim 20$ times the agent diameter for our maze. In this case, the signal-to-noise ratio is given by SNR $\sim \frac{\beta k_0 \sigma}{\gamma_t D_t D_c L_d}$ (or SNR $\sim \frac{B^*}{D^* L_d^*}$ in dimensionless units), which is also independent of the distance to the maze exit. For our chosen system parameters for comparing with the chemical source-seeking strategy in Fig. 1, SNR $\approx 10$. For the parameters used to numerically simulate our experiments with droplet swimmers (i.e., Fig. 3), SNR $\approx 32$. In both cases, the SNR is greater than 1, highlighting the efficiency of the strategy. In our experiments (i.e., Fig. 3, with $\approx 4$ mm-sized PSSNa droplets), the polymer release rate, and consequently the chemotactic drift velocity, decays over time. Based on experimental measurements (see Fig. S2), we model this decay as $k_s(t) = k_0 e^{-t/\tau_d}$, with a characteristic timescale $\tau_d \approx 750$ s. Since the SNR scales linearly with the production rate, it follows the same temporal decay. Starting from an initial value of $SNR \approx 32$, the ratio would drop to $SNR \approx 12$ after $t = \tau_d = 750$ s. Given that the mean exit time for the droplets is significantly shorter ($\approx 321$ s), the SNR remains well above the threshold ($SNR \gg 1$) throughout the entire navigation process, ensuring robust decision-making despite the decaying velocity.

*Estimation of parameters for comparing chemical echolocation and source-seeking strategies*

For the comparison between the strategies, i.e., the simulations corresponding to Fig. 1, we use the parameters from the experiments of [1], where oil-filled droplets with a diameter $\sigma \approx 100$ µm and persistence time (inverse of rotational diffusion coefficient) $\tau_p \approx 20$ s solved a maze by moving up an imposed surfactant gradient. The translational diffusion coefficient of the particle was estimated from an effective Stokes-Einstein relation for spherical particles $D_t = \sigma^2/(3\tau_p) \approx 10^{-10}$ m$^2$ s$^{-1}$. We estimated the droplet mass using $m = \frac{4\pi\rho}{3}(\frac{\sigma}{2})^3 \approx 5 \times 10^{-10}$ kg, where the density of the droplet is $\rho = 10^3$ kg m$^{-3}$. The translation diffusion coefficient of the agent was estimated using $\gamma_t = 6\pi\eta(\sigma/2) \approx 10^{-6}$ kg s$^{-1}$, where $\eta = 10^{-3}$ kg m$^{-1}$ s$^{-1}$ is the dynamic viscosity of water.

The chemical production rate was estimated using $k_0 = \frac{3N_s|dr_d/dt|r_d^2}{N_A r_{fm}^3} \approx 5 \times 10^{-14}$ mol s$^{-1}$, where $N_s = 25$ is the number of surfactant monomers per filled micelle, $r_{fm} = 2.5$ nm is the micellar radius, $r_d = \sigma/2$ is the droplet radius, $|dr_d/dt| = 0.01$ µm s$^{-1}$ is the droplet solubilization rate, and $N_A = 6 \times 10^{23}$ mol$^{-1}$ is the Avogadro constant (see Appendix B.1 of [2]).

The diffusion coefficient of the filled micelles is $\sim 10^2$ µm$^2$ s$^{-1}$ [2]. However, due to the low diffusivity, the advection of the surfactants is dominant for such Marangoni stress-driven droplets and swimmers, and it has been shown that the effect of this advection can be modeled with an enhanced diffusion coefficient [3–5]. Following [3], we estimate the effective chemical diffusion coefficient using $D_c = \sigma u_0$, where $u_0$ is the typical swimmer speed, roughly one body length (i.e., diameter $\sigma$) per second for these active droplets. This gives $D_c \approx 10^4$ µm$^2$ s$^{-1}$.

Using these values, we estimated the units of length, time, and concentration as $r_u \approx 10^{-4}$ m, $t_u \approx 10^2$ s, and $c_u \approx 10^{-4}$ mol m$^{-2}$, respectively, and the dimensionless parameters: $D^* \approx 10^2$, $M^* \approx 10^{-5}$. Finally, to estimate $\beta$ (or equivalently the dimensionless parameter $B^*$), we designed a maze of the same geometry and size as in [1], and performed simulations placing a source at the exit to match the exit times ($\sim 5$ min) of those droplets, which gave us $B^* \approx 2 \times 10^4$.

For the simulations corresponding to all the panels in Fig. 1, we ignored the inertia term in the equation of motion, since $M^* \approx 10^{-5}$ and used the parameter values $B^* = 2 \times 10^4$ and $D^* = 10^2$.

*Estimation of parameters for comparison between simulations and experiments*

To estimate the persistence time of droplet swimmers [6], we recorded the mean squared displacements (MSD) of 13 droplets vs. time (Fig S1) by performing experiments in big tanks of dimension 110 cm × 60 cm. The droplets were released in the center of the tank, and their motion was recorded by a GoPro camera at a speed of 30 frames per second before they reached the boundary. For an active droplet with self-propulsion speed $v_0$, we expect ballistic dynamics, i.e. MSD $\sim 2v_0^2 t^2$, in the intermediate timescale ($t < \tau_p$) and active diffusion for $t > \tau_p$ where MSD $\sim 2v_0^2 \tau_p t$ [7]. We observed the same for our droplets and estimated the persistence time of each trajectory from the fits of MSD vs. time (an example fit of one trajectory is shown in Fig S1). We estimated the average persistence time from all the trajectories as $\tau_p = 3.41 \pm 0.03$ s.

Our droplets are larger compared to the oil-filled droplets of [1], having a diameter $\sigma = 4$ mm. Using the expressions described in the previous section, we estimated $D_t \approx 10^{-6}$ m$^2$ s$^{-1}$, $m \approx 5 \times 10^{-5}$ kg, $\gamma_t \approx 4 \times 10^{-5}$ kg s$^{-1}$. The

production rate is $k_0 \approx 10^{-11}$ mol s$^{-1}$ (see Supplemental Material of [6]). We account for the decay in polymer production by considering an exponentially decaying production term: $k_s(t) = k_0 e^{-t/\tau_d}$, and estimated the decay timescale $\tau_d \approx 750$ s from our experiments (see Fig S2). Using the expression for effective chemical diffusion coefficient in the previous section, we estimated $D_c = \sigma u_0 \approx 2 \times 10^8$ µm$^2$ s$^{-1}$, where the swimmer speed $u_0$ is 15 times the body length, i.e., the diameter $\sigma$ [6] (this is due to the high Peclét number $\approx 10^7$ of our droplets).

Using these values, we estimated the units of length, time, and concentration as $r_u \approx 4 \times 10^{-3}$ m, $t_u \approx 10$ s, and $c_u \approx 6 \times 10^{-7}$ mol m$^{-2}$, respectively, and the dimensionless parameters: $D^* = 2 \times 10^2$, $M^* = 10^{-1}$, $\tau^* = \tau_d D_t/\sigma^2 = 7 \times 10^1$ and $B^* = -8 \times 10^4$ (estimated from the value that closely matches the mean exit time and success fraction in experiments; see Fig 3e in the main text).

*Calculation of the exit time from a point in the maze*

For each trajectory, we calculated the distance from every point along the trajectory to the fixed exit point using a Uniform Cost Search (UCS) (a variant of Dijkstra's algorithm, which employs a priority queue to explore paths in order of increasing distance) [8] pathfinding algorithm. For each point in the trajectory, the algorithm determined the shortest path (and its length, i.e., the distance) to the exit and noted the time taken to reach the exit from that point. This provided us with a set of exit times corresponding to the distances of each point from the exit. To determine the mean exit time for a given distance, we first aggregated all the exit times across all the trajectories corresponding to that distance. We then calculated the mean and standard deviation of these aggregated exit times.

*Calculation of success fraction in simulations*

The success fraction (in Fig. 3e) quantifies how closely the agent trajectories follow the 'ideal' shortest-path-following trajectory. To calculate this, we first used the UCS algorithm to generate this ideal trajectory between the fixed start and exit points in our maze. The idea then is to quantify the degree of overlap of an agent's trajectory with this ideal trajectory. For that, we calculated the Euclidean distance of each point in an agent's trajectory to the nearest point on the ideal trajectory. If the distance falls within a predefined tolerance value (here, 4 agent diameters), we classify the time spent at a point as 'success', otherwise as 'failure' (see Fig. S3). Dividing the total 'success' time (i.e. summed across all points in a given trajectory) by the trajectory's total exit time yields the success fraction for that trajectory.

*Chemical echolocation in different experimental setups*

To assess the general applicability of the chemical echolocation mechanism beyond the specific droplet system used in this work, we analyze the operational criteria derived from our theoretical model. As derived previously, effective echolocation requires that the agent's response time ($t_{\text{response}}$) is shorter than the echo timescale ($t_{\text{echo}}$). This yields the condition $D_c \lesssim \sqrt{k_0 |\beta|/\gamma_t}$. To facilitate comparison with existing literature where the intrinsic chemotactic sensitivity $\beta$ is rarely quantified, we express this criterion in terms of the readily observable chemotactic advection speed (which is often similar to the self-propulsion speed) $u_0$. We note that the chemotactic advection speed can be estimated by $u_0 \approx \frac{|\beta|}{\gamma_t}|\nabla c|$ and substituting the characteristic self-generated gradient derived previously ($|\nabla c| \approx \frac{k_0}{D_c \sigma}$), we obtain the scaling $u_0 \approx \frac{|\beta| k_0}{\gamma_t D_c \sigma}$. Consequently, the feasibility condition reformulates into a comparison between the intrinsic bare chemical diffusivity ($D_c$) and the agent's advective strength ($\sigma u_0$): $D_c \lesssim \sigma u_0$. Physically, this represents a Péclet number condition ($Pe = \sigma u_0/D_c \gtrsim 1$). It implies that for the "echo" to be detected before dissipating, the agent's self-generated advective transport must be comparable to or dominant over the molecular diffusion of the signaling chemicals. Table I lists the feasibility of chemical echolocation in a few different setups based on these criteria.

*Downscaling*

For very small agent sizes (i.e., small agent diameter $\sigma$), their dynamics are overdamped; hence, the only two relevant dimensionless parameters controlling their dynamics are the reduced chemical diffusion coefficient $D^* = D_c/D_t$, and the reduced chemotactic sensitivity $B^* = \beta k_0/\gamma_t D_t^2$. The translational diffusion coefficient $D_t \sim \sigma^{-1}$, drag coefficient $\gamma_t \sim \sigma$, and the production strength $k_0 \sim \sigma^2$ (i.e., proportional to the surface area of the agent). The chemotactic

Table I. Feasibility of chemical echolocation in different setups. Comparison of bare chemical diffusivity ($D_c$) vs. advective strength ($\sigma u_0$). Echolocation is feasible if $\sigma u_0 \gtrsim D_c$ (i.e., high Péclet number regime).

| Setup | Size ($\sigma$) [m] | Speed ($u_0$) [m s$^{-1}$] | Advective strength ($\sigma u_0$) [m$^2$ s$^{-1}$] | Bare diffusivity ($D_c$) [m$^2$ s$^{-1}$] | Expected feasibility ($\sigma u_0 \gtrsim D_c$) |
|---|---|---|---|---|---|
| PSS droplets [this work] | $\sim 4 \times 10^{-3}$ | $\sim 10^{-2}$ | $\sim 4 \times 10^{-5}$ | $\sim 10^{-11}$ (Polymers) | Very high |
| Oil-in-water droplets [9] | $\sim 10^{-4}$ | $\sim 10^{-4}$ | $\sim 10^{-8}$ | $\sim 10^{-10}$ (Micelles) | Very high |
| Rolled-up Ti/Fe/Pt microtubes [10, 11] | $\sim 5 \times 10^{-5}$ | $\sim 2 \times 10^{-4}$ | $\sim 1 \times 10^{-8}$ | $\sim 10^{-9}$ (Dissolved $O_2$) | High |
| Polyaniline/Pt microtubes [12] | $\sim 8 \times 10^{-6}$ | $\sim 3 \times 10^{-3}$ | $\sim 2 \times 10^{-8}$ | $\sim 10^{-9}$ (Dissolved $O_2$) | High |
| Ga/Zn Bubble Motors [13] | $\sim 9 \times 10^{-6}$ | $\sim 4 \times 10^{-4}$ | $\sim 4 \times 10^{-9}$ | $\sim 4 \times 10^{-9}$ (Dissolved $H_2$) | Moderate |
| Janus particles [14–16] | $\sim 0.5 - 10 \times 10^{-6}$ | $\sim 2 - 40 \times 10^{-6}$ | $\sim 2 \times 10^{-11}$ | $\sim 10^{-10} - 10^{-9}$ ($H_2O_2$) | Moderate |

sensitivity parameter scales as $\beta \sim \sigma$ for autophoretic Janus colloids [17], and $\beta \sim \sigma^2$ for droplets [18]. Hence, the dimensionless coefficients depend on size as $D^* \sim \sigma$, and $B^* \sim \sigma^4$ (for autophoretic Janus colloids) or $B^* \sim \sigma^5$ (for droplets). For successful chemical echolocation, one should have $D_t \ll D_c \lesssim \sqrt{k_0|\beta|/\gamma_t}$ or $1 \ll D^* \lesssim \sqrt{B^*}$ in dimensionless units. These criteria are met for droplets of sizes $\sigma \sim 10^2$ µm when keeping all other parameters fixed at the values used in Fig. 3 of the main text. Note that for our droplets, the chemical Peclét number (Pe $\approx 10^7$) is far above the threshold required for spontaneous symmetry breaking to start the motion, consistent with the diffusivity criterion stated above. We performed experiments with smaller droplets of sizes 1 mm and 500 µm, and both these droplets took correct decisions at most maze junctions (see Fig. S4), demonstrating successful chemical echolocation. The same criteria are met for autophoretic Janus colloids of size 1 µm with typical velocities of 10 µm s$^{-1}$ (and $D_c \sim 10^2$ µm$^2$ s$^{-1}$, $D_t \sim 10^{-1}$ µm$^2$ s$^{-1}$ [7]), showing that chemical echolocation is successful down to the microscale. To make chemo-hydrodynamic echolocation effective for smaller droplets and active colloids, one has to reduce the agent's drag coefficient by changing the medium or employ agents that produce chemicals at much higher rates or exhibit enhanced chemical sensitivity. An example of the former is $1-10$ µm Janus colloids moving in Argon gas, primarily by a photophoretic mechanism, where the drag coefficient $\gamma_t$ can be as low as $10^{-12}$ kg s$^{-1}$ [19], i.e., four and six orders of magnitude lower than for autophoretic colloids and droplet swimmers. To some extent, the chemical production rate and chemical sensitivity can be tuned, in turn, by surface functionalization or by using different types of agents, perhaps based on artificial cells and other life-like system components in the future.

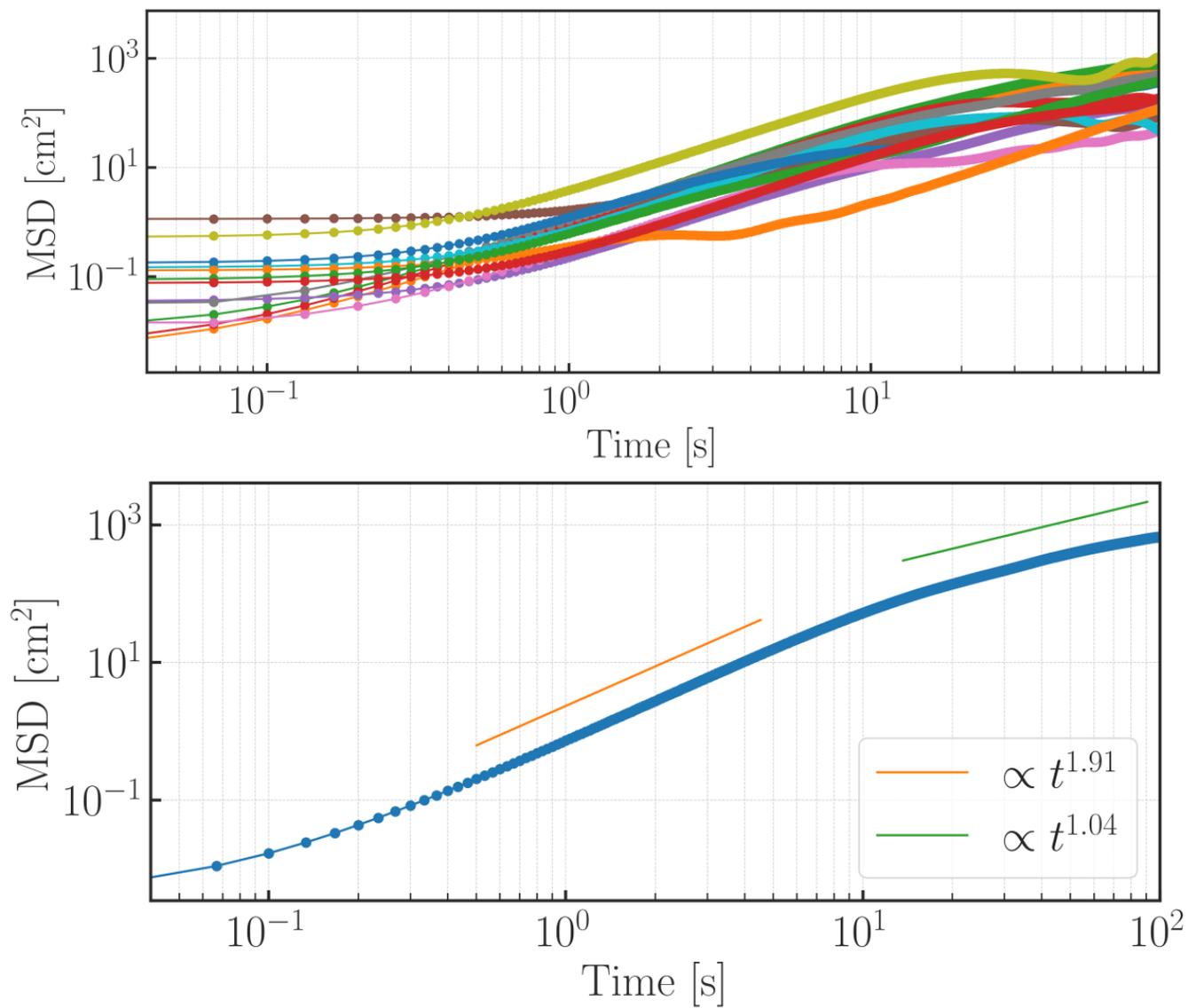

Figure S1. (upper panel) Mean squared displacements (MSD) of different droplet trajectories. (lower panel) Fits showing intermediate ballistic and long-time diffusive dynamics of one exemplary trajectory.

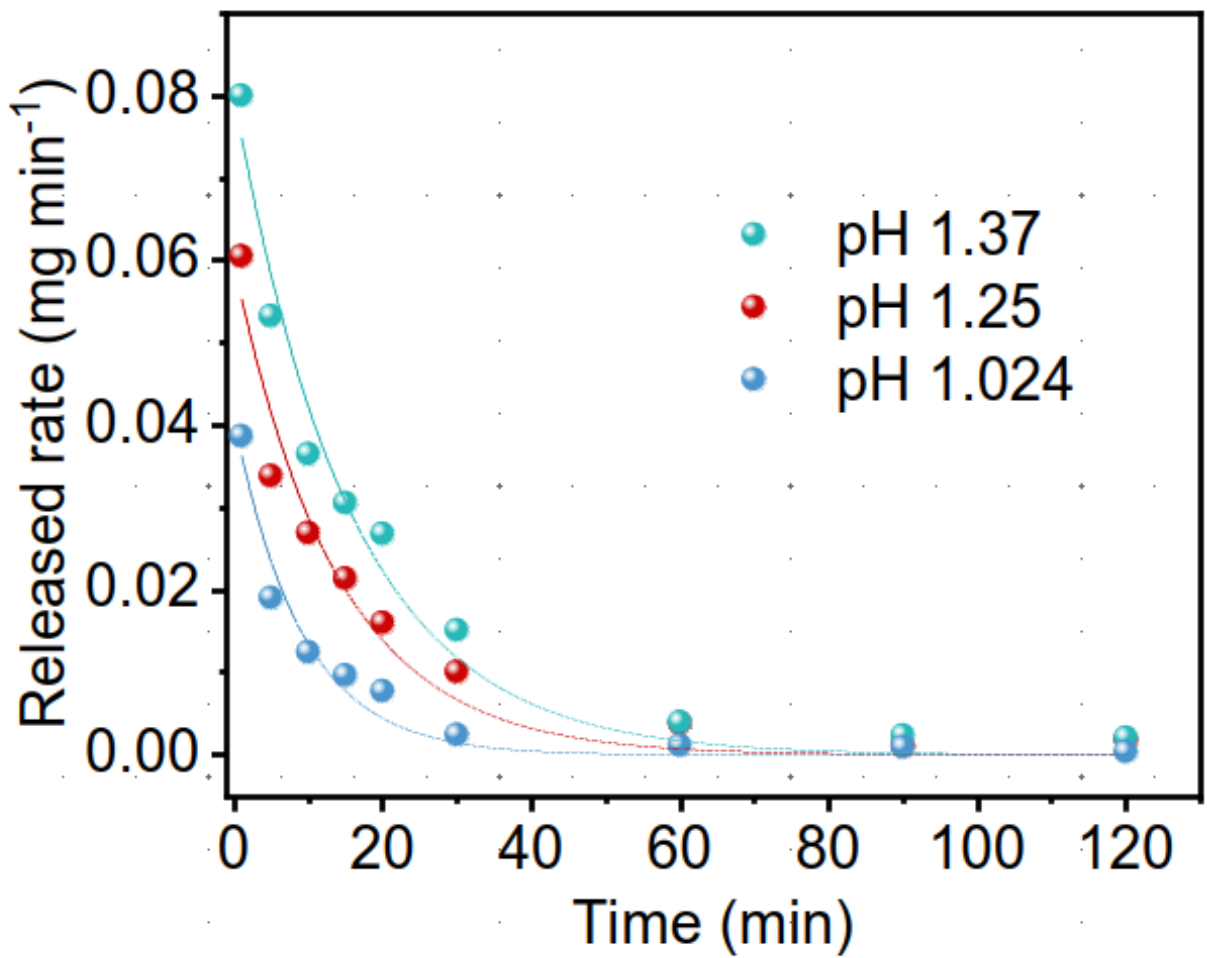

Figure S2. Time evolution of the release rate of PSS by the droplet for different pH. The release rate is fitted to $ae^{-bt}$ (and the released mass is fitted to $\int_0^t ae^{-bt'} dt'$), and the solid lines correspond to the fits. The fits yield: a=0.07987 mg min$^{-1}$; b=0.0638 min$^{-1}$ (pH 1.37), a=0.05954 mg min$^{-1}$; b=0.0727 min$^{-1}$ (pH 1.25), a=0.04061 mg min$^{-1}$; b=0.11048 min$^{-1}$ (pH 1.024).

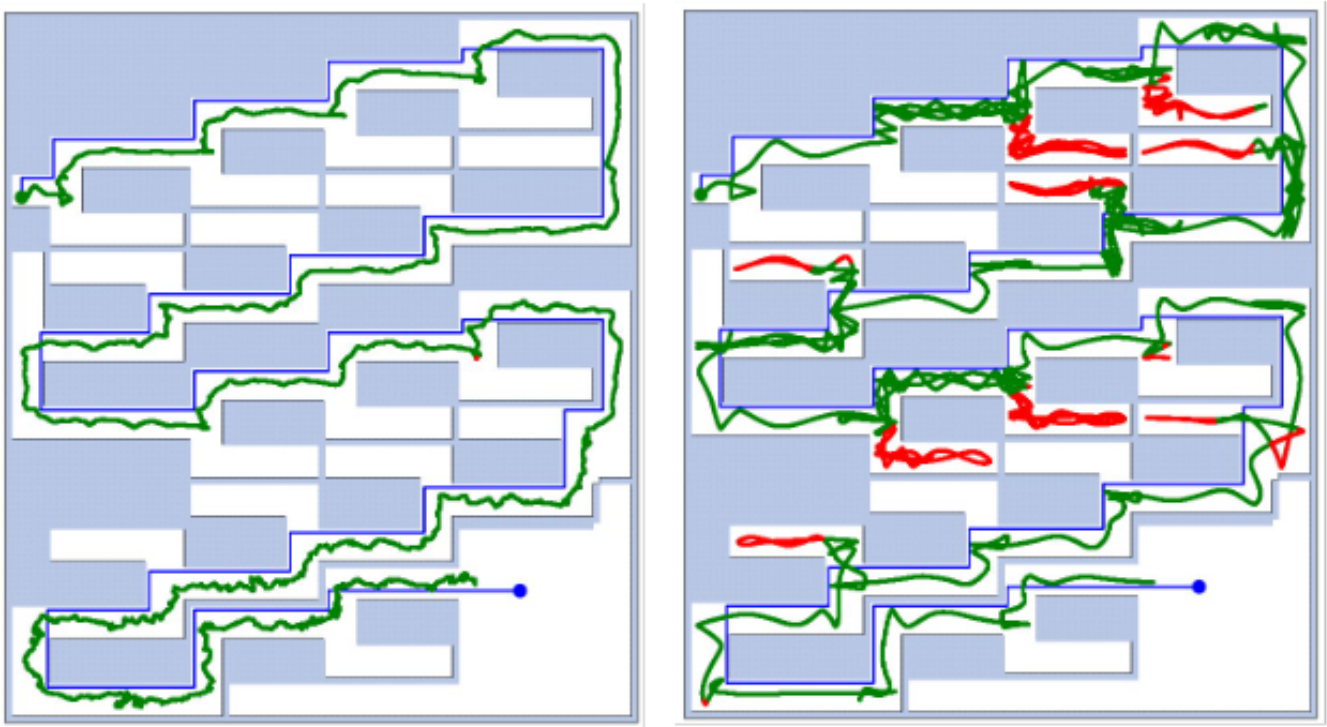

Figure S3. (Left panel) A trajectory with a high success fraction, where the simulated agent's path (green segments) closely follows the ideal Uniform Cost Search (UCS) trajectory (blue line). (Right panel) A trajectory with a lower success fraction, showing more deviation from the ideal path, with red segments indicating points exceeding the 4-agent-diameter distance tolerance.

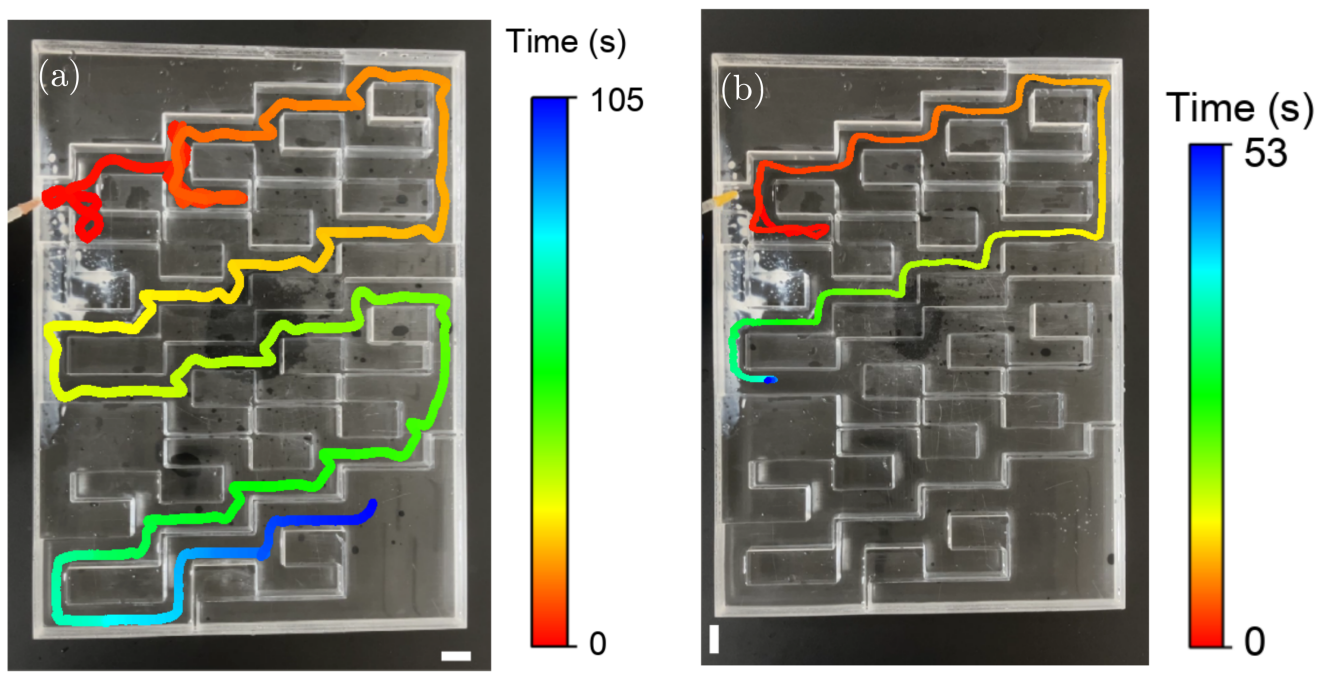

Figure S4. Representative trajectories of small droplets with diameters (a) 1 mm and (b) 500 µm from experiments showing successful decision making at maze junctions. The 500 µm, being smaller, runs out of fuel (PSS) and hence stops before reaching the exit.

**Experiment section**

*Recording droplet swimmer trajectories*

In experiments, we first filled the maze (length: 6.0 cm; width: 2.0 cm; height: 3.0 cm) with an acidic solution of pH 1.25. A droplet of the PEI/PSS solution was deposited at one entrance of the maze from a fixed height of $\approx 1$ cm above the air-water interface. To control the size of the droplet, needles of different diameters were used, i.e., 34G for droplet of diameter 500µm; 34G for droplet of diameter 1 mm; Pipettes for droplet of diameter 4 mm. Then, the droplet starts self-propulsion at the air-water interface as it slowly solidifies from the outer edge to the inside. The videos were recorded with a digital camera at a frame rate of 30 Hz, and the position of the droplet in each frame was obtained by extracting the perimeter using a self-written Python script, from which the trajectory and velocity of the droplet swimmer were calculated.

*Statistical analysis*

The motion from the entrance of the maze to the buffer region is the completion of an experiment. To ensure the reliability and consistency of the results, the data of SF and TTE were presented as "mean ± standard errors" resulting from the average of multiple replicate experimental measurements (n = 60). Standard errors were calculated by dividing the standard deviation by the square root of the sample size.

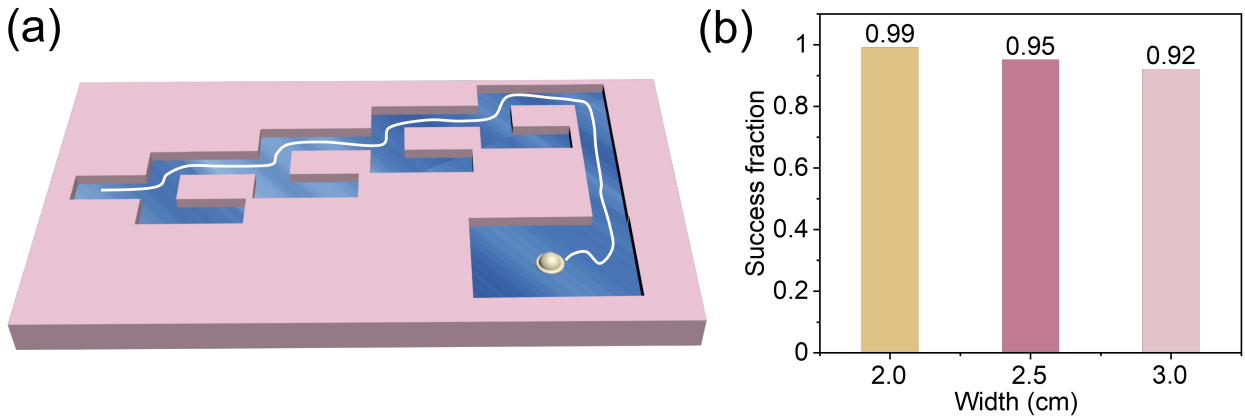

Figure S5. (a) Schematic of a small maze with four junctions. (b) Influence of maze width on the success rate.

*Experiments with smaller maze*

We also performed experiments with a smaller maze containing 4 junctions and a buffer region (Fig. S5) to optimize the geometry of the maze of high success fraction. All the channels have the same length.

*Experiments with varying starting positions of droplets*

We conducted experiments in which droplets were placed at different initial positions within the maze (Fig. S6(a-e)). In all cases, the droplets successfully navigated to the exit. The performance is more reliable when the starting point is not too far from the maze entrance.

*Experiments with multiple droplets*

We performed experiments with multiple droplets placed simultaneously at different initial positions within the maze (Fig. S6f). The droplets generally navigated successfully to the exit, demonstrating robust decision-making, although they occasionally made incorrect turns at junctions. Future investigations will be necessary to understand the underlying interactions in such multi-droplet systems, which affect their collective navigation performance.

*Measurement of flow fields generated by droplet*

The flow field generated by the droplet in the maze was quantified by Particle Image Velocimetry (PIV). Carbon black nanopowders were added, which were uniformly distributed on the air-water interface before the droplet was dipped into the acidic water. Videos were recorded at a frame rate of 30 Hz, from which the flow field was analyzed. As shown in (Fig. S7), there is a backflow echo from the dead end, which pushes the droplet to move to the open end. This further confirms the chemo-hydrodynamic echolocation mechanism.

**Influence of Maze Geometry on Navigation Performance**

*Effect of Dead-arm length.* To test the limits of the echolocation mechanism, we simulated agents in a series of Y-junctions where the length of the dead-end arm ($L_d$) was systematically increased relative to the open arm ($L_o$). As shown in Fig. S9a, the success fraction is highest ($\sim 1.0$) when the dead arm is short ($L_d/L_o < 0.5$). Here, the chemical "echo" builds up rapidly in the dead arm and creates a strong repulsive gradient towards the exit at the junction. As the dead arm becomes longer, the time required for the signal to reflect back to the junction increases,

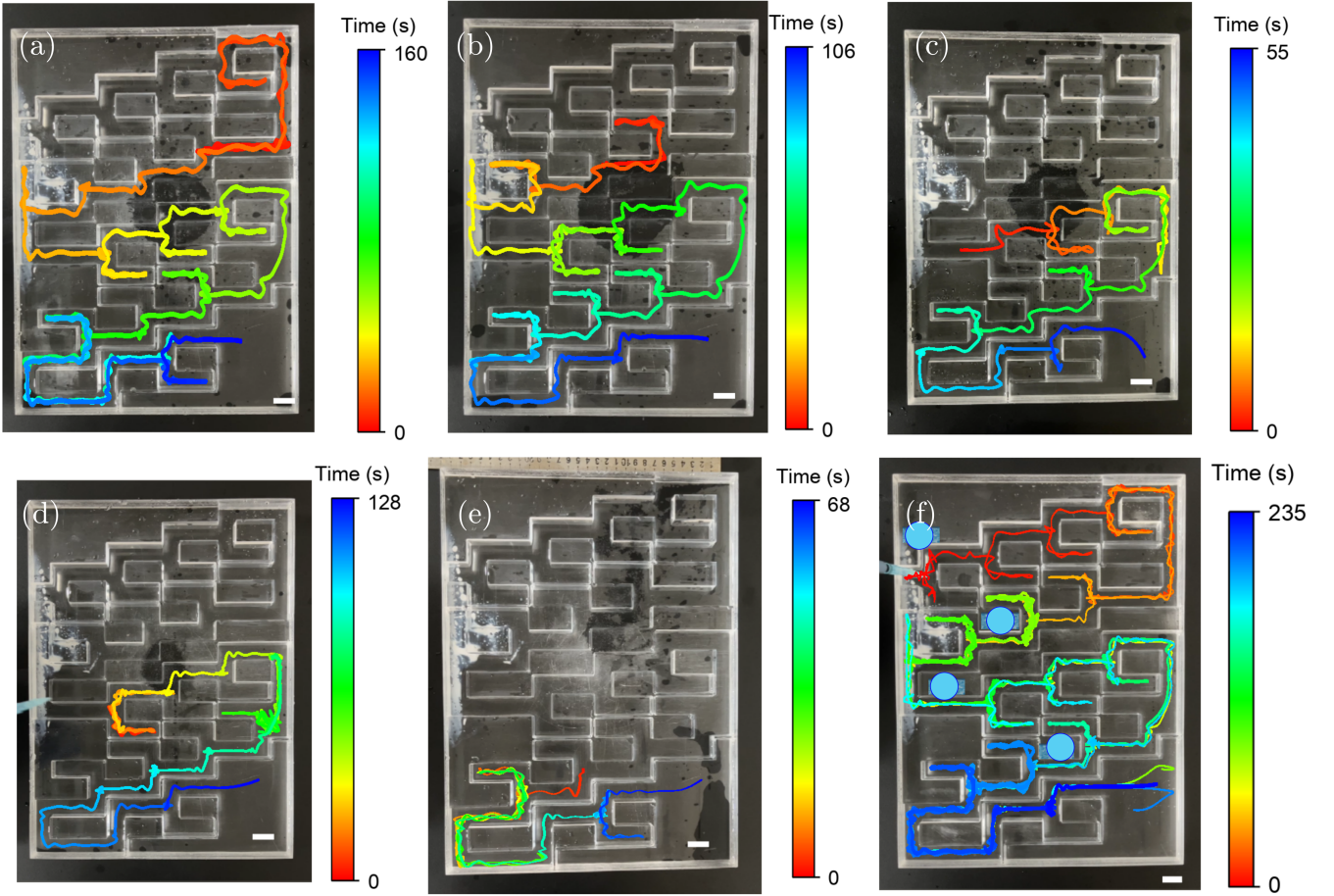

Figure S6. (a–e) Individual experimental trajectories of a single droplet launched from different starting points within the maze, highlighting consistent decision-making at junctions and successful avoidance of dead ends. (f) Overlaid trajectories of four droplets introduced simultaneously at distinct positions (marked by cyan-filled circles) successfully reaching the exit.

hence the steepness of the gradient decreases at the same time. Consequently, the average success fraction drops but remains robust ($\approx 0.6$) even when the dead arm is twice as long as the open path ($L_d/L_o = 2.0$). This shows that while short branches facilitate faster decisions, the echolocation mechanism remains effective for extended dead ends. We also investigated the impact of the available space within the maze for chemical accumulation by scaling the maze dimensions, effectively increasing the dead-arm area relative to the agent (Fig. S9b). We find that the decision-making reliability persists (success fraction $\gtrsim 0.65$) even when the dead-arm area is nearly 2000 times larger than the agent, demonstrating that the mechanism is not limited to geometries with very small dead-arms.

We experimentally complemented these findings by varying the droplet diameter ($\sigma = 4, 2$, and $0.5$ mm) within a fixed maze geometry (channel width $W = 2$ mm). That way, we effectively varied the ratio of the dead-arm area to the agent's area (Fig. S10a). For the largest droplets ($\sigma = 4$ mm), the relative dead-arm volume is small, facilitating rapid chemical buildup and a strong repulsive echo, resulting in a high success fraction ($\approx 90\%$, see Fig. S10b). Conversely, for the smallest droplets ($\sigma = 500$ $\mu$m), the dead arms represent a significantly larger relative volume (analogous to the large-area regime in the simulations). Despite this increased available space, which delays signal accumulation, the droplets successfully navigate the maze with a robust success rate of $\approx 70\%$, experimentally confirming that the mechanism is effective even when the relative dead-end dimensions are large. We further validated these findings experimentally by observing droplets navigate a maze featuring multiple dead ends with lengths 1.5–2.0 times that of the open paths (Fig. S10c). In ten parallel trials, the droplets consistently identified the correct path, achieving a success fraction of 100% and confirming the robustness of the echolocation mechanism even against extended dead branches.

*Effect of maze channel width.* We experimentally investigated the role of channel width of the maze by observing the trajectories of droplets with different diameters ($\sigma = 0.5, 2, 4$ mm) in mazes of varying channel widths ($W$). This

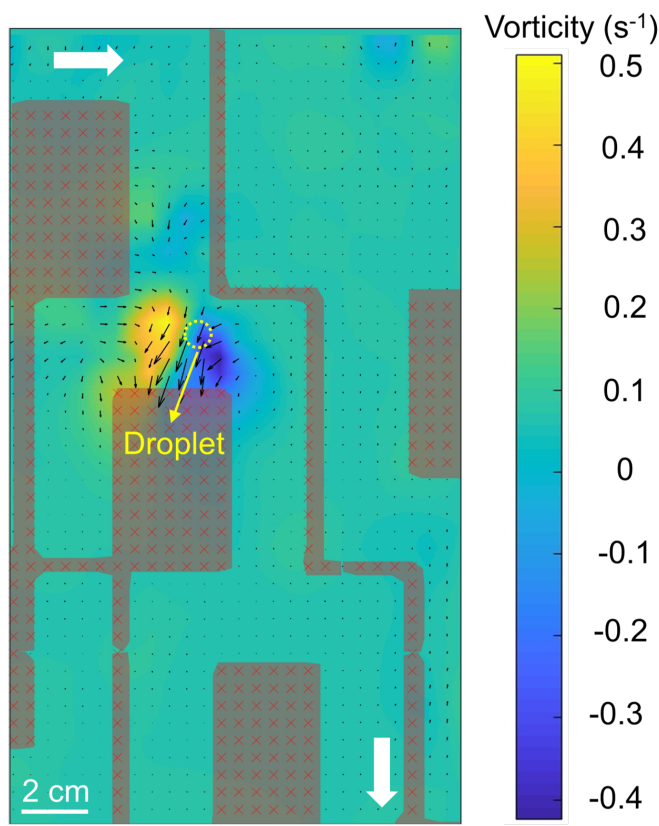

Figure S7. Flow field (black arrows) generated by the droplet (yellow-bordered disc) at the junction, showing signatures of chemo-hydrodynamic echo. The gray regions are walls.

provided a range of width-to-diameter ratios ($W/\sigma$) from roughly 5 to 50. Representative trajectories are shown in Fig. S11. The droplet can navigate in mazes when the width/diameter ratio is smaller than 50. At $W/\sigma \gtrsim 50$, the droplet performs chiral motion within the channel as reported in our previous work for droplets in a Petri dish [6], which greatly hinders its maze solving.

### List of movies

1. **Solving a maze by automated decision-making using chemical echolocation.** Animated version of Fig. 1a showing the simulated trajectory of an agent performing automated decision-making to solve a maze using chemical echolocation without any external sources.

2. **Solving a maze using a chemical source-seeking strategy.** Animated version of Fig. 1b showing the simulated trajectory of an agent solving a maze by sensing a chemical gradient created by an external source at the maze exit.

3. **Experimental realization of automated decision-making in droplet swimmers.** Experimental trajectory of a droplet swimmer (corresponding to Fig. 3a) solving a maze using chemical echolocation.

4. **Collective navigation of multiple droplet swimmers.** Experimental trajectories of multiple droplets released simultaneously into the maze.

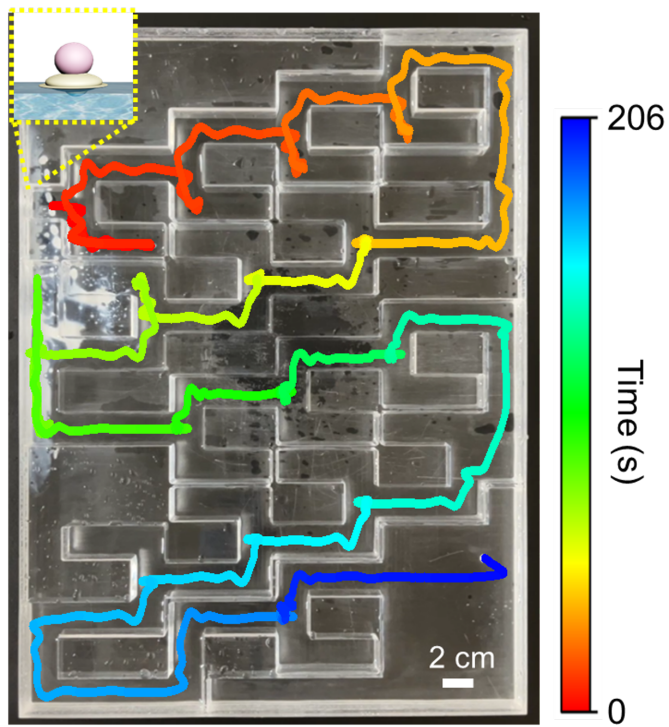

Figure S8. Typical trajectory of a droplet (diameter 4 mm) successfully carrying a big foam of diameter 3 mm to the maze exit.

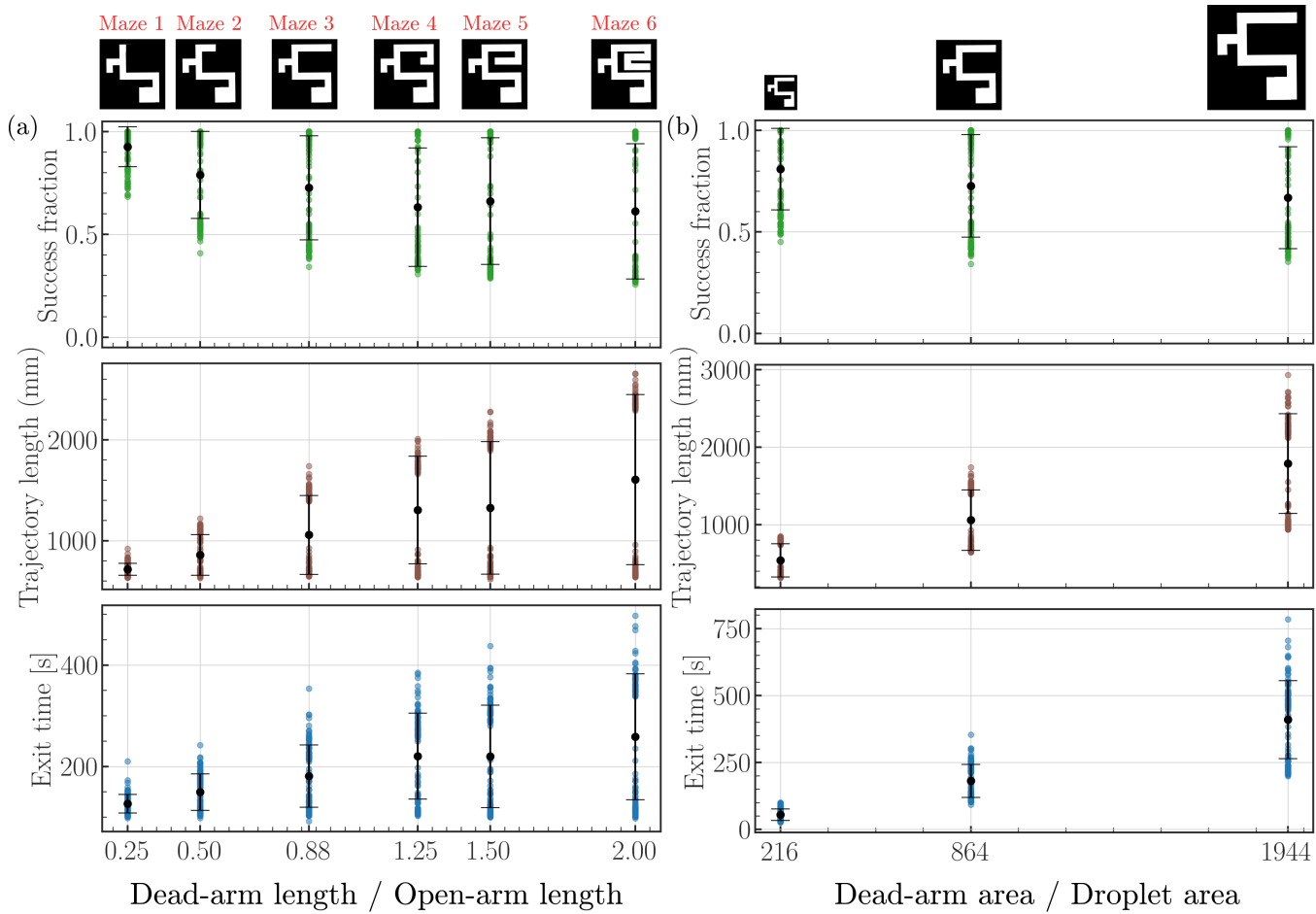

Figure S9. **Influence of dead-arm length and area on chemical echolocation.** (a) Success fraction, trajectory length, and exit time plotted as a function of the ratio between the dead-end length and the open-arm length. The data corresponds to simulations in mazes 1–6 (top panel), where the dead-end length is systematically increased. (b) The same performance metrics for Maze 3 scaled by factors of 1, 2, and 3 (representing a 1x, 2x, and 3x increase in maze dimensions relative to the agent), shown as a function of the ratio of the dead-arm area to the droplet size (area). Black dots indicate mean values with standard deviation error bars; colored points represent individual simulation runs.

14

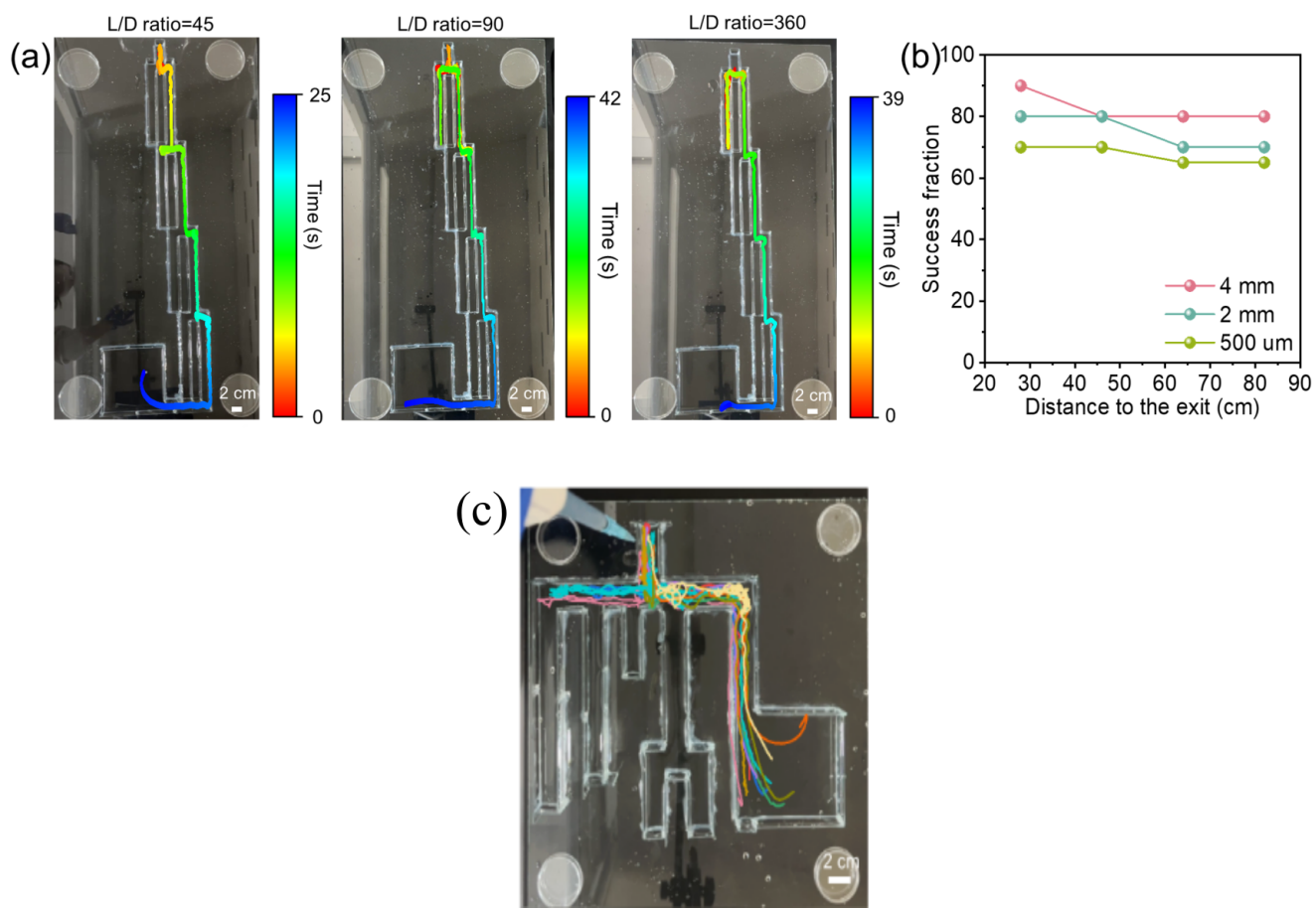

Figure S10. (a) Typical trajectories of droplets of various diameters (from left to right) from experiments: 4 mm, 2 mm and 0.5 mm in a maze of length 18 cm and width 2 mm. (b) Success fraction as a function of distance to the exit for droplets of various diameters. (c) Trajectories of droplets navigating a maze of multiple dead ends of various lengths.

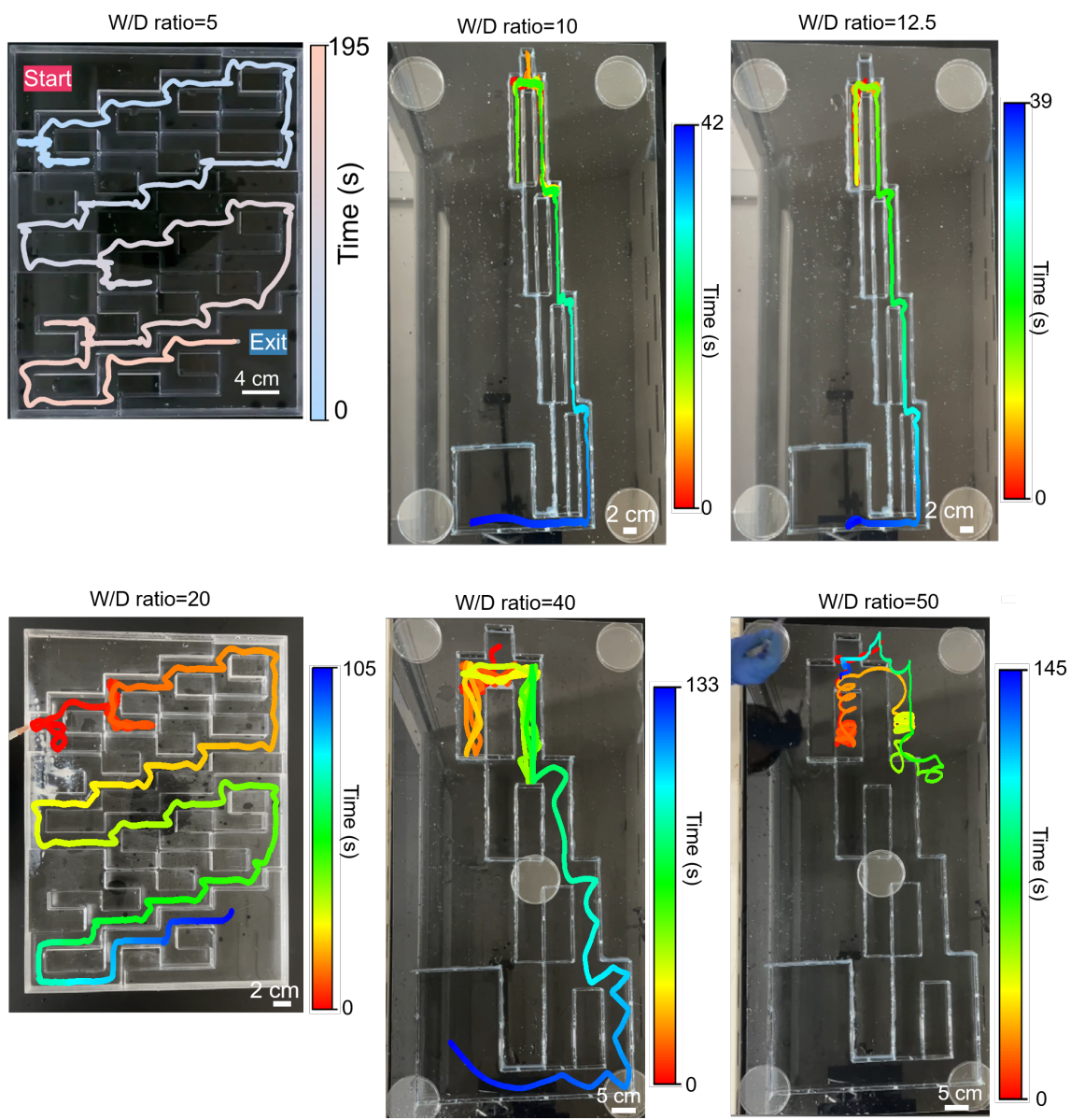

Figure S11. Typical trajectories of droplets of various diameters (from left to right) from experiments: 4 mm, 2 mm and 0.5 mm in maze of various widths.


\fi

\end{document}